\documentclass[pdflatex,sn-mathphys-num]{sn-jnl}

\usepackage{graphicx}%
\usepackage{multirow}%
\usepackage{amsmath,amssymb,amsfonts}%
\usepackage{amsthm}%
\usepackage[title]{appendix}%
\usepackage{xcolor}%
\usepackage{textcomp}%
\usepackage{manyfoot}%
\usepackage{booktabs}%
\usepackage{algorithm}%
\usepackage{algorithmicx}%
\usepackage{algpseudocode}%
\usepackage{enumerate}%
\usepackage{thmtools}%

\theoremstyle{thmstyleone}%
\newtheorem{theorem}{Theorem}%
\newtheorem{proposition}[theorem]{Proposition}%
\newtheorem{lemma}[theorem]{Lemma}%
\newtheorem{corollary}[theorem]{Corollary}%
\newtheorem{observation}[theorem]{Observation}%

\theoremstyle{thmstylethree}%
\newtheorem{definition}[theorem]{Definition}%
\newtheorem{problem}{Problem}%

\theoremstyle{thmstyletwo}%
\newtheorem{remark}[theorem]{Remark}%

\newif\ifshownotes
\shownotesfalse

\newcommand{\MinTCover}{\textsc{$k$-MinTimelineCover}}

\newcommand{\oneMinTCover}{\textsc{$1$-\MinTCover{}}}
\newcommand{\OPT}{\mathrm{OPT}}
\newcommand{\ALG}{\mathrm{ALG}}
\newcommand{\NC}{\mathrm{NC}}

\usepackage{tikz}
\usetikzlibrary{decorations.pathreplacing,positioning, arrows.meta,shapes}
\usetikzlibrary{patterns}

\tikzset{main node/.style={circle,draw,minimum size=0.2cm,fill=black!100,inner sep=0pt},
            }

\tikzset{example node/.style={circle,draw,minimum size=0.4cm,inner sep=3pt},
            }

\tikzset{forbidden node/.style={circle,fill=blue!50,draw,minimum size=0.4cm,inner sep=3pt},}

\begin{document}

\title[Covering the timeline in temporal graphs]{Heuristics for covering the timeline in temporal graphs}

\author[1]{\fnm{Riccardo} \sur{Dondi}}\email{riccardo.dondi@unibg.it}

\author*[2]{\fnm{Rares-Ioan} \sur{Mateiu}}\email{rares.mateiu@s.unibuc.ro}

\author[2]{\fnm{Alexandru} \sur{Popa}}\email{alexandru.popa@fmi.unibuc.ro}

\affil[1]{\orgname{Universit\`a degli Studi di Bergamo},
  \orgaddress{\city{Bergamo}, \country{Italy}}}

\affil*[2]{\orgdiv{Department of Computer Science},
  \orgname{University of Bucharest},
  \orgaddress{\city{Bucharest}, \country{Romania}}}

\maketitle

\begin{abstract}

We consider Minimum Timeline Cover (\MinTCover{}), a variant of the Vertex Cover problem on temporal graphs, introduced for the summarization of complex interaction data. Given a temporal graph $G$, \MinTCover{} asks for at most $k$ activity intervals for each vertex, such that
each temporal edge has at least an endpoint active at its timestamp, with the objective of minimizing the sum of interval lengths.
The problem is NP-hard, and for $k > 1$ it is hard to approximate within any factor, since deciding whether a solution of span $0$ exists is already NP-complete. 
While the literature has mainly focused on the cases $k=1$, in this contribution we also consider the case $k > 1$. We first present an ILP formulation that is able to solve the problem on moderate size instances. 
We then show that, although no approximation factor is attainable in general, a $(T-1)$-approximation is achievable for every $k$ on the instances whose snapshots are matchings, via a minimum-cost flow formulation of the underlying assignment problem; for $k = 1$ this improves the best known factor of $4(T-1)$. Next, we design a polynomial-time heuristic based on local search, built on top of an existing method, which applies to larger temporal graphs. Finally, we extend the learning-based approach of Lazzarinetti et al.~\cite{DBLP:conf/time/LazzarinettiMZD24,lazzarinetti2025dlmintc+} from $k = 1$ to an arbitrary $k$.

An experimental evaluation against proven optima shows that, on synthetic graphs with social structure, the learned method has a mean empirical ratio of about $1.14$ to the optimum and reduces the total span by more than $40\%$ with respect to the strongest published heuristic, also on instances with up to $5{,}000$ vertices, where the exact solver no longer finds good solutions within two minutes; on two real contact networks, without retraining, its advantage is smaller and depends on the network.
\

\end{abstract}

\section{Introduction} 
\label{introduction}

Complex systems are usually modeled through graphs or networks in order
to understand their properties and behaviors.
The availability of temporal information of relations between vertices
has led to the introduction of temporal graphs~\cite{ref_1,ref_2,DBLP:journals/jcss/KempeKK02}, where 
edges, called temporal edges, represent interactions between  two entities at a given time.
Note that different variations of the temporal graph model have been
introduced; here we consider a model where the time domain, over which
the temporal edges are defined, is discrete \cite{ref_1,ref_2,DBLP:journals/jcss/KempeKK02}.

Temporal graphs have been considered in several domains, 
e.g., social network analysis \cite{tang2009temporal}, computational biology~\cite{DBLP:journals/netmahib/HosseinzadehCGD23} and epidemiology \cite{DBLP:journals/iandc/DeligkasDEGS24}.
We consider an approach defined for the summarizations of temporal interactions in social networks \cite{ref_3,ref_6,ref_9,ref_9bis}. In this context,
a user activity in a social network (called activity timeline) 
is represented with one or more time intervals; the goal is to 
represent users interaction, so that if an interaction
is observed at time $t$, then at least one
of the user activity timeline includes $t$~\cite{ref_9bis,ref_9,ref_10,ref_11,ref_12,ref_B,ref_D}. 
The objective function, following a parsimonious approach, is the 
minimization of  the overall length (called \emph{span}) of timeline activities or the minimization of the maximum length of the timeline activity. Here, we focus on the first objective function.

Informally, the problem we study can be stated as follows. We are given a
log of timestamped interactions between entities, 
where every interaction must be explained by defining, for each entity, a small number of time
intervals during which it was active. An interaction (represented by the edges) observed at time $t$ is explained when at least one of its two participants (represented by the vertices) is active at $t$.
Each entity is allowed at most $k$ such intervals, and the cost of an interval is its duration. 
\MinTCover{} asks for the activity intervals of minimum total
duration that explain all the observed interactions. A formal definition
is given in Section~\ref{preliminaries}.

Several papers have studied the complexity of 
\MinTCover{}~\cite{ref_9,ref_10,ref_11,ref_12},
also in some restricted variant. In particular, 
the case $k=1$, 
where the activity timeline of each vertex is defined as a single interval, has been deeply studied, due also to the link with a
foundational problem in theoretical computer science, that is
\textsc{Minimum Vertex Cover}.
For $k=1$ \MinTCover{} is NP-hard and hard to approximate within
constant factor~\cite{ref_9,ref_10,ref_11,ref_12}.
For $k \geq 2$, the problem is even harder, as even deciding
if there exists a solution of span $0$ is an NP-complete problem~\cite{ref_9}.

Due to the hardness of the problem and of its restrictions, some heuristic methods have been developed, for the general problem~\cite{ref_9,ref_9bis}
and for the restriction with $k=1$ \cite{ref_9,ref_9bis,DBLP:conf/time/LazzarinettiMZD24}.
Only the  paper that has introduced \MinTCover{} has designed practical methods for the problem \cite{ref_9} when $k \geq 2$, thus in this contribution we focus on designing methods for it. In this contribution we study both the approximability of the problem and practical methods for it, for every $k$. First, we design an ILP method that is able to solve the problem for moderate size instances, then we present a polynomial-time heuristic  for \MinTCover{}, based on local search, so that it is     applicable even on larger temporal graphs. Second, we show that,
although no approximation factor is attainable in general for $k \ge 2$, a
deterministic $(T-1)$-approximation exists for every $k$ when every snapshot is a matching; we complement it with a hardness result and a
resource-augmentation guarantee (Section~\ref{sec:theory-k}). Third, we extend the learning-based approach of~\cite{lazzarinetti2025dlmintc+} from $k=1$ to the budgets $k \le 10$ seen in training, through a decoder that assigns each temporal edge to one of its endpoints and then reconstructs the timeline of each vertex exactly (Sections~\ref{sec:heuristic} and~\ref{sec: ML}). We evaluate the learned method against the exact optimum and against the heuristics of~\cite{ref_9}, on synthetic temporal graphs with social structure and on a real contact network (Section~\ref{sec:experiments}). A local-search improvement
step for the heuristics of~\cite{ref_9} was studied in the conference version of this paper~\cite{mateiu2025time} and it is not among the contributions evaluated here.

\subsection{Related Works}

The complexity of \MinTCover{}, including the parameterized and approximation complexity, has been considerably investigated. 
\MinTCover{} is NP-hard, and remains so even under restricted variants. For $k=1$, the problem is NP-hard when at most one temporal edge is defined in each timestamp \cite{ref_12}, when the time domain consists of three timestamps and every vertex has degree at most two in each snapshot \cite{ref_12}, and when the time domain consists of only two timestamps \cite{ref_10}. In this paper we show that the first of these restrictions remains NP-hard for every fixed $k$ (Theorem~\ref{thm:hardness}).

The case $k \geq 2$ is harder, in a way that changes the kind of
guarantee one can hope for. Rozenshtein et al.~\cite{ref_9} have shown
that for every $k \geq 2$ it is already NP-complete to decide whether an
instance admits a solution of span \emph{zero}; the same question was
later studied for $k=2$ in \cite{ref_10}. Since an approximation
algorithm with factor $c$ must return a solution of cost at most
$c \cdot 0 = 0$ on such an instance, and a cost is never negative, any
approximation algorithm would decide this NP-complete problem exactly.
Hence for $k \geq 2$ the problem admits no polynomial-time approximation
within any factor, however large, unless $\mathrm{P}=\mathrm{NP}$.

\MinTCover{} have also been considered through the lens of parameterized complexity, for parameters of the number of vertices,  the number of timestamps in the time domain, and $k$ \cite{ref_10}. 
For $k=1$, \MinTCover{} has been shown to be fixed-parameter tractable, for the parameter the span of a solution,  when the time domain consists of two timestamps \cite{ref_10} and later for any number of timestamps \cite{ref_11}. It is unlikely that the previous result can be extended to larger values of $k$ because of the previous statement regarding the NP completeness of deciding on the existence of a size 0 solution\cite{ref_10}. 
The parameterized complexity has been also studied 
considering structural parameters, for example the degree, the vertex-interval-membership-width and interval-membership-width
\cite{DBLP:conf/iwpec/HerrmannKMS25,DBLP:journals/tcs/HerrmannKMS26}.

As for the approximation complexity,
we have already noted that \MinTCover{} cannot be approximated
within any factor. For $k=1$, while the problem is known to be hard to approximate within constant factor assuming the Unique Game Conjecture \cite{ref_10,ref_11}, it admits a $O(T \log |V|)$-approximation algorithm~\cite{ref_B,ref_D} ($T$ is the number of timestamps of the time domain, $V$ is the set of 
vertices).

As for heuristics, Rozenshtein et al. \cite{ref_9} have presented \textbf{k-Inner}, which  is based on solving a problem, called  \textsc{k-Coalesce}, that requires to cover some predetermined timestamps. These timestamps represent an estimation of the activity timeline of a vertex.
In our contribution, we present a local search heuristic which is meant to be built on top of this heuristic, in order possibly improve the computed solutions (that is we are able to reduce the span).

We note that other complexity results and  a heuristic, called \textbf{k-Budget}, have been given for the problem that
has as objective function the minimization of the maximum span \cite{ref_9,ref_10}.

Other heuristics have been proposed recently for the case 
$k=1$ by Lazzarinetti et al. \cite{DBLP:conf/time/LazzarinettiMZD24,lazzarinetti2025dlmintc+}, who proposed \textit{DLMinTC+}, the first learning-based method for the problem. Rather than constructing a cover by local search, they train a neural model to predict activity intervals directly. Extending the learning-based approach to arbitrary $k$ is one of our contributions in this paper.

In Section \ref{sec:theory-k} we study the approximability of the problem for $k > 1$ and snapshots are matchings, and we give an approximation algorithm for this case, together with a hardness result and a resource-augmentation guarantee. In Section \ref{sec:heuristic} we describe the exact per-vertex reconstruction shared by our algorithms, and in Section \ref{sec: ML} a learning-based method. In Section \ref{sec:experiments} we give an experimental evaluation of our methods on synthetic and real-world datasets. We conclude the paper with some future directions described in Section \ref{conclusion}.

\section{Preliminaries} 
\label{preliminaries}

A temporal graph $G = (V,E)$, with $V$ a set of vertices and $E$ a set of temporal edges, which are defined on time domain $[T] = [1,T]$ 
of timestamps (note that in the model we consider the time domain is discrete and $T$ represents the maximum timestamp value).
Each temporal edge is represented by a timestamped triplet $(u,v,t)$ such that $u,v \in V$ and $t \in [T]$ represents the time when the interaction  between $u$ and $v$ occurred. 
Note that in the temporal graphs we consider, temporal edges are undirected and unweighted and we write $n = |V|$ for the number of vertices and $m = |E|$ for the number of temporal edges of $G$.

For $t \in [T]$, the snapshot $G_t = (V,E_t)$ is the static graph layout at time $t$ on the full vertex set with $E_t = \{(u,v) : (u,v,t) \in E\}$. We write $N_t(v)$ for the neighbor set of $v$ in $G_t$ and $deg(v,t) = |N_t(v)|$ for its degree at time $t$. 

Consider a vertex $v$ and two timestamps $s_v \in [T]$ and $e_v \in [T]$ with $s_v \le e_v$, $I_v = [s_v, e_v]$ is an activity interval
of $v$.
An \emph{activity timeline} $\mathcal{T}_k(v)$ of a vertex $v \in V$ is a set of at most $k$ (where $k \geq 1$) activity intervals
$I_{v_i} = [s_{v_i}, e_{v_i}], i \in [j] $, with $j \le k$, where 
$s_{v_i} \le e_{v_i}$, with the constraint that $e_{v_i} < s_{v_{i+1}}$ for each $i \in [j-1] $. The last constraint guarantees
that the activity intervals of an activity timeline of a vertex
are time disjoint; the timeline may also be empty ($j=0$).Since timestamps are integers, two intervals of a timeline may occupy consecutive timestamps: for instance $[t,t]$ and $[t+1,t+1]$ are two
disjoint intervals, and $v$ is active at both $t$ and $t+1$. This is
consistent with the interpretation of timestamps as the instants at which
interactions are recorded: the vertex is not required to be active
between two consecutive instants. Given a vertex $v \in V$ and an activity timeline $\mathcal{T}_k(v)$ of $v$, we say that $v$ is \emph{active} in a timestamp $t$ that is included in an interval of $\mathcal{T}_k(v)$; if $t$ is not included in an interval of $\mathcal{T}_k(v)$, $v$ is said to be \emph{inactive} in $t$.

Moreover, we define the \emph{span} of interval $I_{v_i}$ $\delta(I_{v_i}) = e_{v_i} - s_{v_i}$ as the duration of the interval $i$ for the vertex $v$. 
We define the span of an activity timeline  $\mathcal{T}_k(v)$ of $v$, consisting of the intervals $I_{v_i} = [s_{v_i}, e_{v_i}], i \in [j], j \le k $, as
\[
S(\mathcal{T}_k(v)) = \sum_{i=1}^j \delta(I_{v_i}).
\]
with $S(\mathcal{T}_k(v)) = 0$ for an empty timeline.

The activity timeline of temporal graph $G$ is defined as:
\[
\mathcal{T}_k =  \bigcup_{v \in V} \mathcal{T}_k(v)
\]

The \emph{sum span} of the activity timeline $\mathcal{T}_k$ of a graph $G$ is defined as the sum spam over all vertices in the temporal graph:

\begin{equation*}
    S(\mathcal{T}_k) = \sum_{v \in V} S(\mathcal{T}_k(v)) =  \sum_{v \in V} \sum_{j \in [1 \dots k]} \delta (I_{v_i}) 
\end{equation*}

Note that a vertex may use fewer than $k$ activity intervals, or none at all.

Now, we define how an activity timeline covers temporal edges of
a temporal graph.

\begin{definition}
Given a temporal graph $G = (V,E)$ and an activity timeline 
$\mathcal{T}_k$ of $G$, $\mathcal{T}_k$ covers $G$ if $\forall (u,v,t) \in E$, there is an interval $I_v \in \mathcal{T}_k(v) $ or $I_u \in \mathcal{T}_k(u) $ such that $t \in I_v$ or $t \in I_u$.
\end{definition}

Now, we are able to formally define the problem we are interested in.

\begin{problem}$k$-Minimum Timeline Cover (\MinTCover{})\\
\textbf{Input:} A temporal graph $G = (V,E)$, $k \in \mathbb{N}$.\\ 
\textbf{Output:} An activity timeline $\mathcal{T}_k$ that covers $G$ and minimizes the total sum span $S(\mathcal{T}_k)$.
\end{problem}

We denote by $\OPT_k(G)$ the minimum sum span of an activity timeline that covers $G$ with at most $k$ intervals per vertex, that is, the optimum of \MinTCover{} on $G$.
Note that in the following, we will focus on the case $k \geq 2$.

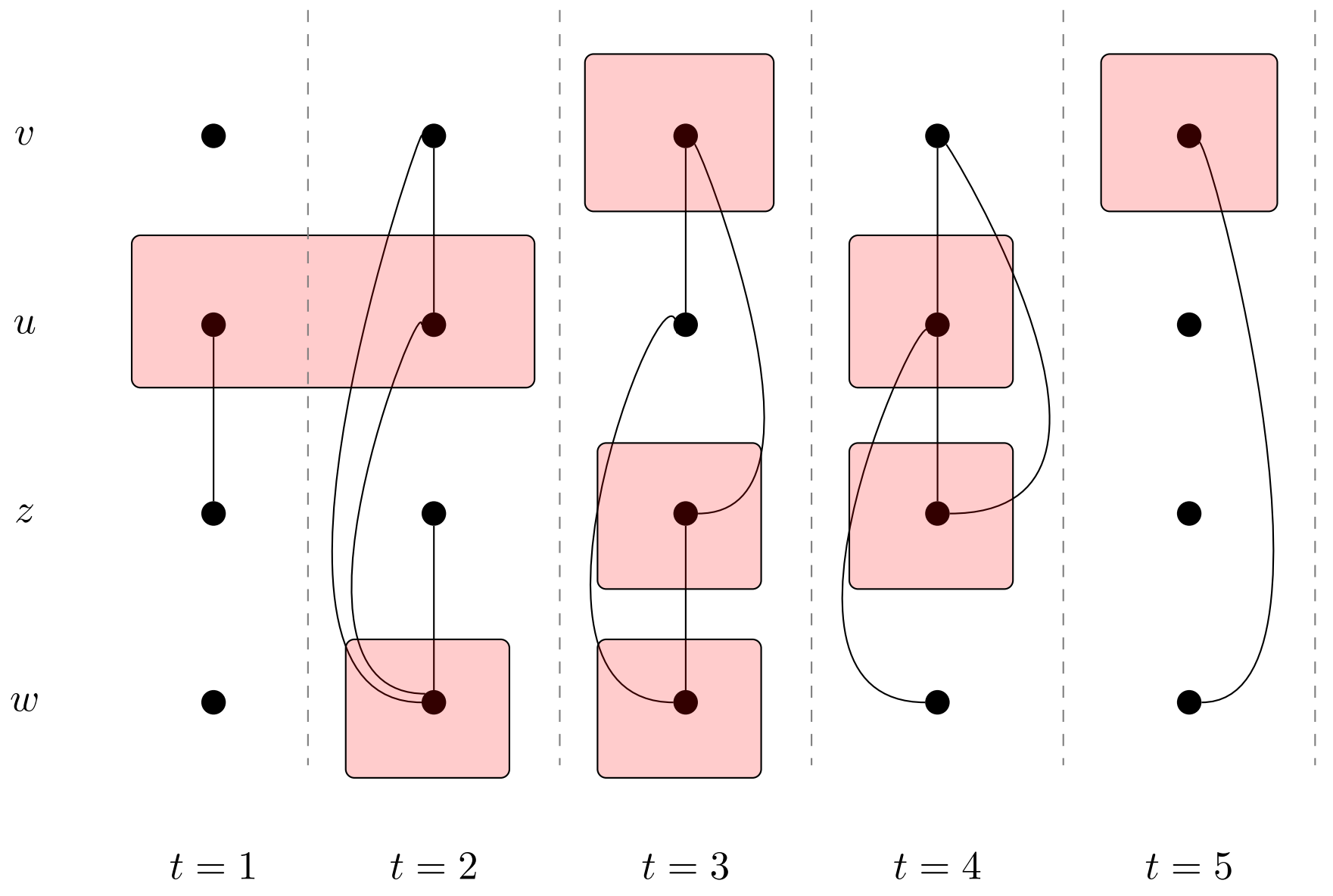
\begin{figure}

  \centering

  \begin{tikzpicture}[scale=1.1]
    \node at (-0.75,0) { $v$};
    \node at (-0.75,-1.5) { $u$};
    \node at (-0.75,-3) { $z$};
    \node at (-0.75,-4.5) { $w$};
               

    \node[main node] (01) at (0.75,0){};

    \node[main node] (02) at (0.75,-1.5) { };
    
    \node[main node] (03) at (0.75,-3){};
    
    \node[main node] (04) at (0.75,-4.5){};

    \draw (02.south) -- (03.north);

    \node[main node] (1) at (2.5,0){};

    \node[main node] (2) at (2.5,-1.5) { };
    
    \node[main node] (3) at (2.5,-3){};
    
    \node[main node] (4) at (2.5,-4.5){};

    \draw (1.south) -- (2.north);
    

    \draw (1.west) .. controls +(down:-1mm) and +(right:-16mm) ..  (4.west);

     \draw (2.west) .. controls +(down:-3mm) and +(right:-13mm) ..  (4.north west);   

    \draw (3.south) -- (4.north);

    \node[main node] (5) at (4.5,0){};
    
    \node[main node] (6) at (4.5,-1.5) { };

    \node[main node] (7) at (4.5,-3){};
    
    \node[main node] (8) at (4.5,-4.5){};

    \draw (5.south) -- (6.north);

    \draw (5.south east) .. controls +(down:-1mm) and +(right:12mm) .. (7.east);

    \draw (6.south west) .. controls +(down:-9mm) and +(right:-15mm) ..  (8.west);

    \draw (7.south) -- (8.north);
    

    \node[main node] (9) at (6.5,0){};
    \node[main node] (10) at (6.5,-1.5) { };
    
    \node[main node] (11) at (6.5,-3){};
    
    \node[main node] (12) at (6.5,-4.5){};

    \draw (9.south) -- (10.north);
    
    \draw (10.south) -- (11.north);

    \draw (9.south east) .. controls +(down:-0.3mm) and +(right:+18mm) ..  (11.east);    

    \draw (10.south west) .. controls +(down:-4mm) and +(right:-15mm) ..  (12.west);



    \node[main node] (13) at (8.5,0){};
    \node[main node] (14) at (8.5,-1.5) { };
    
    \node[main node] (15) at (8.5,-3){};
    
    \node[main node] (16) at (8.5,-4.5){};

    \draw (13.south east) .. controls +(down:-4mm) and +(right:13mm) .. (16.east);


\draw[fill=red, fill opacity=0.2, rounded corners=0.5ex] (3.7,-0.6) rectangle (5.2,0.65);

\draw[fill=red, fill opacity=0.2, rounded corners=0.5ex] (7.8,-0.6) rectangle (9.2,0.65);

\draw[fill=red, fill opacity=0.2, rounded corners=0.5ex] (0.1,-0.79) rectangle (3.3,-2);

\draw[fill=red, fill opacity=0.2, rounded corners=0.5ex] (5.8,-0.79) rectangle (7.1,-2);

\draw[fill=red, fill opacity=0.2, rounded corners=0.5ex] (3.8,-2.44) rectangle (5.1,-3.6);

\draw[fill=red, fill opacity=0.2, rounded corners=0.5ex] (5.8,-2.44) rectangle (7.1,-3.6);

\draw[fill=red, fill opacity=0.2, rounded corners=0.5ex] (1.8,-4) rectangle (3.1,-5.1);

\draw[fill=red, fill opacity=0.2, rounded corners=0.5ex] (3.8,-4) rectangle (5.1,-5.1);

\draw[dashed,thin,gray] (1.5,1)--(1.5,-5);
\draw[dashed,thin,gray] (3.5,1)--(3.5,-5);
\draw[dashed,thin,gray] (5.5,1)--(5.5,-5);
\draw[dashed,thin,gray] (7.5,1)--(7.5,-5);
\draw[dashed,thin,gray] (9.5,1)--(9.5,-5);

\node at (0.75,-5.8){$t=1$};
\node at (2.5,-5.8){$t=2$};
\node at (4.5,-5.8){$t=3$};
\node at (6.5,-5.8){$t=4$};
\node at (8.5,-5.8){$t=5$};

\end{tikzpicture}

\caption{An example of \MinTCover{} with $k=2$. The activity timeline of each vertex is represented with two pink rectangles. Note that the only vertex that has positive span is $u$, all the other vertices
have span equal to $0$.
Every temporal edge is covered by at least an activity timeline.
}
\label{fig:toygraph}
\end{figure}

Figure \ref{fig:toygraph} shows an example of timeline covering over a 2-timestamps graph with 5 vertices.

Considering the problem definition, we can make an important observation: a cover requires only that each temporal edge have one active endpoint, leaving free the choice of which endpoint takes responsibility for it. Making that choice explicit decomposes the problem across vertices.

\begin{definition}[Assignment]
An assignment is a map $a:E\to V$ with $a(u,v,t) \in \{u,v\}$. The required set of a vertex $v$ under $a$ is 
\begin{equation*}
S_v(a)\;=\;\{\,t\in\mathcal{T}\;:\;\exists\,e=(u,v,t)\in E \text{ with } a(e)=v\,\},
\end{equation*}

the set of timestamps at which $a$ makes $v$ responsible for some edge $e$. 
\end{definition}

Based on the assignment definition, we can make the following observation:
\begin{observation}\label{obs:assignment-feasible}
For any assignment $a$, any family of timelines $\{\mathcal{T}_k(v)\}_{v \in V}$ in which each $\mathcal{T}_k(v)$ covers $S_v(a)$ is a feasible solution for k-\textsc{MinTimelineCover}.
\end{observation}

Each edge is assigned to exactly one endpoint, and that endpoint is active at the edge's timestamp; feasibility follows immediately. The significance is that once $a$ is fixed, the vertices become independent: the minimum sum-span achievable under $a$ is the sum over vertices of the minimum span of a $k$-interval cover of $S_v(a)$. The converse also holds, so that
\[
\OPT_k(G) \;=\; \min_{a}\ \sum_{v \in V} \mathrm{minSpan}\bigl(S_v(a),k\bigr),
\]
where $\mathrm{minSpan}$ is defined below: given an optimal covering timeline, assign each temporal edge to an endpoint that is active at its timestamp; the timeline of each vertex $v$ then covers $S_v(a)$, so its span is at least $\mathrm{minSpan}(S_v(a),k)$. Choosing $a$ well is therefore the whole difficulty, a point which we return to in Sections~\ref{sec:theory-k} and~\ref{sec: ML}.

The objective function that measures how expensive a required set is for the
vertex that must cover it repeats throughout the paper, so we name it
here.

\begin{definition}[Minimum span of a set of timestamps]
\label{def:minspan}
Let $S \subseteq [T]$ be a nonempty set of timestamps and let $k \geq 1$. We denote by $\mathrm{minSpan}(S,k)$ the minimum of $\sum_{i=1}^{j} \delta(I_i)$ over all families of $j \leq k$ pairwise time-disjoint intervals $I_1, \dots, I_j$ whose union contains $S$; that is, the span of a cheapest activity timeline of at most $k$ intervals that covers every timestamp of $S$.
\end{definition}

\begin{lemma}\label{lemma:minspan}
Let $S = \{t_1 < \dots < t_p \}$ and let $g_i = t_{i+1} - t_i$, for $i \in [1\dots p-1]$, be its consecutive gaps, and let $g^*_1 \ge g^*_2 \ge \dots \ge g^*_{p-1}$ be the same gaps sorted in nonincreasing order. Then
\begin{equation*}
    \mathrm{minSpan}(S,k)\;=\;(t_p-t_1)\;-\;\sum_{i=1}^{\min(k,p)-1} g^*_i 
\end{equation*}
and in particular we can set $\mathrm{minSpan}(S,k) = 0$ whenever $k \ge p$.
\end{lemma}

\begin{proof}
Any cover of $S$ by $j$ disjoint intervals partitions $S$ into $j$ consecutive blocks and its span is at least the sum over blocks of (last element - first element), which equals $(t_p - t_1)$ minus the $j - 1$ gaps separating the blocks. The span is therefore minimized by choosing the $j - 1$ largest gaps to separate the blocks, and taking $j  = \min(k,p)$ maximizes the number of gaps removed. Conversely the intervals spanning each block achieve this value.


\end{proof}

Alongside the closed form of Lemma~\ref{lemma:minspan} we record two
elementary bounds relating the \emph{number} of timestamps a vertex must
cover to the \emph{span} it must pay for them. They are what allows a
solution to be compared with a relaxation that counts active timestamps
rather than interval lengths, and they are used in
Section~\ref{sec:theory-k}. For the same section we define the \emph{Union Graph}, which is the structure that is created when we merge the timestamps into parallel edges.  
\begin{definition}[Union multigraph]
\label{def:union}
The \emph{union multigraph} of a temporal graph $G = (V,E)$ is the
undirected multigraph $G_U = (V,E_U)$ on the same vertex set that
contains one edge $\{u,v\}$ for every temporal edge $(u,v,t) \in E$.
\end{definition}

\begin{lemma}\label{lemma:bridge}
Let $S \subseteq [T]$ be nonempty with $|S| = p$, and let $k \ge 1$.
Then
\begin{enumerate}[(a)]
  \item $\mathrm{minSpan}(S,k) \;\le\; (T-1)\max(0,\,p-k)$; and
  \item every activity timeline of at most $k$ intervals whose intervals cover $S$ has span at least $p-k$.
\end{enumerate}
\end{lemma}

\begin{proof}

(a) If $p \le k$ then $\mathrm{minSpan}(S,k) = 0$ by
Lemma~\ref{lemma:minspan}. If $p > k$, the same lemma gives
$\mathrm{minSpan}(S,k) = (t_p - t_1) - \sum_{i=1}^{k-1} g^*_i$; since $t_p - t_1$ is the sum of all $p-1$ consecutive gaps of $S$, this is the sum of the $p-k$ smallest gaps, and every gap is at most $T-1$.

(b) An interval $[s,e]$ contains $e-s+1$ timestamps, that is,
$\delta([s,e])+1$. Hence $j \le k$ intervals of total span $\sigma$ contain at most $\sigma + j \le \sigma + k$ timestamps in total, and containing all $p$ elements of $S$ forces $\sigma \ge p-k$.
\end{proof}

\section{An ILP Formulation for \MinTCover{}} 
\label{sec:ILP}

In the following, we present an ILP model formulation for \MinTCover{}. Let $G = (V,E)$ be a temporal graph on the time domain $[T]$, and let $k$ be the maximum number of activity intervals per vertex.
The model uses three families of variables. The binary variable $x_v^t$, $v\in V$, $t \in [T]$, indicates whether vertex $v$ is active at timestamp $t$. The binary variable $y_v^t$, $v \in V$, $t \in [T-1]$, records a change of state of $v$ between $t$ and $t+1$, and is used to bound the number of maximal blocks of consecutive active timestamps of $v$. Finally, the integer variable $z_v \in \{0,\dots,k\}$, $v \in V$, is the number of activity intervals used by $v$.

The variable $z_v$ is needed because the set of timestamps at which a vertex is active does not determine its timeline, nor its span. If $v$ is active at the timestamps $t$ and $t+1$ only, its timeline can be the single interval $[t,t+1]$, of span $1$, or, when $k\ge2$, the two intervals $[t,t]$ and $[t+1,t+1]$, of total span $0$. A formulation that charges each maximal block of consecutive active timestamps as a single interval would therefore overestimate the optimum. The following identity makes the relation between active timestamps, number of intervals and span precise.
For a set $A \subseteq [T]$, a \emph{run} of $A$ is a maximal set of consecutive integers contained in $A$, and $r(A)$ denotes the number of runs of $A$; we set $r(\emptyset)=0$.

\begin{restatable}{theorem}{ilp}\label{thm:ilp}
Let $G = (V,E)$ be a temporal graph on the time domain $[T]$ and let $k \ge 1$. The optimum of the following integer linear program equals $\OPT_k(G)$. Moreover, an optimal activity timeline can be obtained from any optimal solution of the program in $O(|V|\,T)$ time.
\begin{align}
\min\ & \sum_{v\in V}\Bigl(\sum_{t=1}^{T} x_{v}^{t} - z_v\Bigr) \label{eq-1}\\
\text{s.t. }\ & x_{u}^{t} + x_{v}^{t} \ge 1 && \forall (u,v,t) \in E \label{eq-2}\\
& y_{v}^{t} \ge x_{v}^{t} - x_{v}^{t+1},\quad y_{v}^{t} \ge x_{v}^{t+1} - x_{v}^{t} && \forall v \in V,\ t \in [T-1] \label{eq-3}\\
& x_{v}^{1} + \sum_{t=1}^{T-1} y_{v}^{t} + x_{v}^{T} \le 2k && \forall v \in V \label{eq-4}\\
& z_v \le \sum_{t=1}^{T} x_{v}^{t} && \forall v \in V \label{eq-5}\\
& x_v^t \in \{0,1\},\ y_v^t \in \{0,1\},\ z_v \in \{0,1,\dots,k\} \notag
\end{align}
\end{restatable}

We conclude with some remarks on the formulation. The variables $y_v^t$ do not appear in the objective function, so the two lower bounds of constraint~\eqref{eq-3} suffice: a solution may set $y_v^t$ above $|x_v^{t+1}-x_v^t|$, but this only tightens constraint~\eqref{eq-4}. The terms $x_v^1$ and $x_v^T$ in constraint~\eqref{eq-4} account for a run that starts at the first timestamp or ends at the last one, which no variable $y_v^t$ records. The program does not force any vertex to be active: a vertex may take $x_v^t = 0$ for every $t$ and $z_v = 0$, at cost $0$, and the program is feasible for every $T$ and every $k \ge 1$, since activating every vertex at every timestamp gives a single run per vertex.


\section{An approximation algorithm for \MinTCover{}}
\label{sec:theory-k}
Dondi and Popa~\cite{ref_B, ref_D} gave the first approximation
algorithms for the single-interval problem: an $O(T\log n)$-approximation for \textsc{MinTCover} and, for the restriction \textsc{1-MinTCover} in which every timestamp carries at most one temporal edge, a $4(T-1)$-approximation obtained by hitting 
the union graph 
with a feedback vertex set. This section asks which of these guarantees survive the transition to $k$ activity intervals per vertex. The answer is not trivial: for every $k\ge2$, deciding whether a solution of span $0$ exists is NP-complete~\cite{ref_9}, so no polynomial-time algorithm can approximate \MinTCover{} within \emph{any} factor unless $\mathrm{P}=\mathrm{NP}$, and the structural lemmas that
drive the $k=1$ analysis fail outright for $k\ge2$. Our main positive result is that on instances whose snapshots are
matchings --- a class containing every \textsc{1-MinTCover} instance --- the relaxation used in~\cite{ref_B} can be solved
\emph{exactly} by a minimum-cost flow for every $k$, which yields a deterministic $(T-1)$-approximation for all $k\ge1$ and, at $k=1$, improves the $4(T-1)$ bound. We complement this with NP-hardness of the restricted problem for every $k$ and
a resource-augmentation guarantee for the general problem. We need to remind that throughout our paper the spans use the following duration convention: an interval $[l,r]$ costs $r-l$ and a single active timestamp is free.

Beyond Section~\ref{preliminaries} we use the following. The
\emph{global degree} $\Delta(v)$ of a vertex is the number of temporal edges incident to $v$ over the whole time domain, and
$\Delta_L(G)=\max_{v\in V,\,t\in[T]}deg(v,t)$ is the maximum local
degree. We call $G$ a \emph{matching-snapshot} instance if
$\Delta_L(G)\le1$, that is, every snapshot $G_t$ is a matching. The \emph{union multigraph} $G_U$ has vertex set $V$ and one edge $(u,v)$ per temporal edge $(u,v,t)\in E$; distinct timestamps on the same pair therefore give parallel edges. Finally, \oneMinTCover{} denotes the restriction of \MinTCover{} to instances with at most one temporal edge per timestamp, the generalisation to an arbitrary $k$ of the restriction studied in~\cite{ref_B}; every \oneMinTCover{} instance is a matching-snapshot instance.

\subsection{The approximability boundary for $k\ge2$}
\label{sec:theory-boundary}

\begin{proposition}\label{prop:barrier}(\cite{ref_9})
For every $k\ge2$, \MinTCover{} admits no polynomial-time approximation algorithm with any approximation factor, unless
$\mathrm{P}=\mathrm{NP}$.
\end{proposition}

The next statement locates the source of this hardness: it is the possibility of two simultaneous interactions at one vertex, not the interval budget. Recall that a multigraph is \emph{$k$-orientable} if
its edges can be oriented with in-degree at most $k$ everywhere; by Hakimi's theorem~\cite{hakimi1965degrees} this holds if and only if every vertex set $S$ induces at most $k|S|$ edges, and an orientation is
computable in polynomial time by flow techniques.

\begin{proposition}\label{prop:span0}
For a matching-snapshot instance $G$, $\OPT_k(G)=0$ if and only if $G_U$ is $k$-orientable. In particular span-$0$ feasibility is decidable in
polynomial time on this class, for every $k$.
\end{proposition}

\begin{proof}
Given an orientation with in-degree at most $k$, let each vertex cover the edges oriented towards it by one interval of span $0$ each: at most $k$ intervals per vertex, total span $0$. Conversely, in a timeline of span $0$ every interval is a single timestamp; assign each temporal edge to an endpoint one of whose intervals is its timestamp. Since
$\Delta_L(G)\le1$, a vertex receives at most one edge per interval, hence at most $k$ edges in total, and the assignment is an orientation of $G_U$ with in-degree at most $k$.
\end{proof}

For $k=1$ the algorithms of~\cite{ref_B} rest on the fact that an instance admitting a solution of span $s$ has, after preprocessing, at
most $2s$ vertices of global degree at least $3$, whose removal leaves disjoint paths and hence a feedback vertex set of size at most $2s$. One
might hope that for general $k$ the vertices of global degree at least $2k+1$ play this role. They do not.

\begin{restatable}{proposition}{nodegree}\label{prop:no-degree}
For every $k\ge2$ and every $h\in\mathbb{N}$ there is a connected \oneMinTCover{} instance with $\OPT_k\le2$, minimum global degree at
least $k+1$, no connected component coverable with span $0$, and more than $h$ vertices of global degree at least $2k+1$.
\end{restatable}

Hence for $k\ge2$ neither degrees nor cycles capture the structure that
a bounded-span solution leaves behind. Proposition~\ref{prop:span0}
indicates the replacement: the residue at span $0$ is a $k$-orientable
multigraph, so it is assignments, not degrees, that must be optimised.
Doing so gives an algorithm that is both stronger and simpler than the
one it generalises.

\subsection{A $(T-1)$-approximation on matching snapshots}
\label{sec:theory-flow}

Following~\cite{ref_B}, we relax the requirement that a vertex be active on \emph{intervals}. In the non-consecutive relaxation $\NC_k$, every vertex $v$ receives a set $S_v\subseteq[T]$ of active
timestamps such that every temporal edge $(u,v,t)$ has
$t\in S_u\cup S_v$, at cost $\sum_{v\in V}\max(0,|S_v|-k)$; write $\NC_k(G)$ for the minimum cost. Applying Lemma~\ref{lemma:bridge}(b) to
the timestamps covered by each vertex of an optimal timeline gives
\begin{equation}\label{eq:nc-lower}
\NC_k(G)\;\le\;\OPT_k(G).
\end{equation}
The relaxation is exactly the assignment view of
Section~\ref{preliminaries}, freed from the interval structure:

\begin{lemma}\label{lem:assignment-form}
For every temporal graph $G$ and every $k\ge1$,
\[
\NC_k(G)\;=\;\min_{a}\ \sum_{v\in V}\max\bigl(0,\,|S_v(a)|-k\bigr),
\]
the minimum being over all assignments $a:E\to V$.
\end{lemma}

\begin{proof}
For any assignment $a$ the sets $S_v(a)$ are feasible for the
relaxation, since every edge $e=(u,v,t)$ has $t\in S_{a(e)}(a)$ and
$a(e)\in\{u,v\}$; this gives $\le$. Conversely, given feasible
$(S_v)_{v\in V}$, define $a$ by assigning each edge to an endpoint
active at its timestamp. Then $S_v(a)\subseteq S_v$ for every $v$, so
$|S_v(a)|\le|S_v|$ and the cost does not increase, which gives $\ge$.
\end{proof}

Minimising over assignments is NP-hard in general --- its span-$0$
question is that of \MinTCover{} --- but on matching snapshots it becomes
a flow problem, because there each assigned edge contributes its own
timestamp.

\begin{lemma}\label{lem:flow}
If $\Delta_L(G)\le1$ then an assignment attaining the minimum of
Lemma~\ref{lem:assignment-form}, and hence the value $\NC_k(G)$, is
computable in polynomial time by minimum-cost flow.
\end{lemma}

\begin{proof}
Since $\Delta_L(G)\le1$, distinct edges assigned to $v$ carry distinct
timestamps, so $|S_v(a)|=|a^{-1}(v)|$ and the objective of
Lemma~\ref{lem:assignment-form} counts assigned edges. Build the network
with a source supplying $m$ units, one node per temporal edge receiving
one unit and sending it, at capacity $1$ and cost $0$, to either of its
two endpoint nodes, and, from each vertex node to the sink, one arc of
capacity $k$ and cost $0$ together with one arc of capacity $m$ and cost
$1$. Integral minimum-cost flows of value $m$ exist and are computable
in polynomial time~\cite{ahuja1988network}; they correspond to assignments, and
since the per-vertex cost is convex an optimal flow routes the first $k$
units of each vertex at cost $0$ and every further unit at cost $1$,
so its cost is $\sum_v\max(0,|a^{-1}(v)|-k)$.
\end{proof}

\begin{theorem}\label{thm:flow}
On matching-snapshot instances of \MinTCover{}, hence on all instances of
\oneMinTCover{}, the following deterministic polynomial-time algorithm is
a $(T-1)$-approximation for every $k\ge1$: compute a minimum-cost
assignment $a$ by Lemma~\ref{lem:flow}, and give every vertex $v$ a
timeline $\mathcal{T}_k(v)$ of minimum span covering $S_v(a)$, of span
$\mathrm{minSpan}(S_v(a),k)$. Its span satisfies
\[
\ALG\;\le\;(T-1)\,\NC_k(G)\;\le\;(T-1)\,\OPT_k(G),
\]
and $\ALG=0$ whenever $\OPT_k(G)=0$.
\end{theorem}

\begin{proof}
The output is feasible by Observation~\ref{obs:assignment-feasible}. By
Lemma~\ref{lemma:bridge}(a) and the optimality of $a$
(Lemmas~\ref{lem:assignment-form} and~\ref{lem:flow}),
\[
\ALG=\sum_{v\in V}\mathrm{minSpan}(S_v(a),k)
\;\le\;(T-1)\sum_{v\in V}\max\bigl(0,|S_v(a)|-k\bigr)
\;=\;(T-1)\,\NC_k(G),
\]
and~\eqref{eq:nc-lower} completes the chain. If $\OPT_k(G)=0$ then
$\NC_k(G)=0$, so $|S_v(a)|\le k$ for every $v$ and every interval
returned has span $0$.
\end{proof}

\begin{corollary}\label{cor:k1}
For $k=1$, the restriction of \MinTCover{} to one temporal edge per
timestamp admits a deterministic polynomial-time
$(T-1)$-approximation, improving the $4(T-1)$-approximation
of~\cite{ref_B}.
\end{corollary}

\begin{remark}\label{rem:flow-discussion}
At $k=1$ the algorithm replaces the preprocessing, the subdivided union
graph, the $2$-approximate feedback vertex set and the two inequalities
$\ALG\le2(T-1)|F^*|$ and $|F^*|\le2\,\OPT$ of~\cite{ref_B} by a
single exact flow computation: the two factors of $2$ vanish because the flow optimises over all span-$0$ residues ($1$-orientable multigraphs) rather than over forests, and compares the relaxation value to $\OPT$ directly. The factor $T-1$ is the worst-case price of one merged interval per unit of relaxation cost, and instances attaining
$\Theta(T)$ exist. 
In practice, from the experiments of Section \ref{sec:experiments} the algorithm is far from this bound: on random \oneMinTCover{} instances with $k\in\{1,2,3\}$ and proven ILP optima, the certified chain $\NC_k\le\OPT_k\le\ALG\le(T-1)\NC_k$ held on every instance and the ratio $\ALG/\OPT_k$ averaged $3.2$. Two further consequences are worth recording. First, $\NC_k(G)$ is a polynomial-time computable \emph{lower bound} on $\OPT_k(G)$, available on instances far beyond the reach of the exact solver and hence usable to benchmark heuristics there. Second, the algorithm is precisely the decoding pipeline of Section~\ref{sec: ML} with the learned assignment replaced by the flow assignment, the per-vertex step being the same $\mathrm{minSpan}$ computation; the learned and the provable decoders therefore differ only in how $a$ is chosen. On instances with $\Delta_L(G)\ge2$ the algorithm still returns a feasible timeline of span at most $(T-1)$ times the flow value, but the flow value is then no longer a lower bound on $\OPT_k$, as Proposition~\ref{prop:barrier} requires.
\end{remark}

\subsection{NP-hardness of \oneMinTCover{} for every $k$}
\label{sec:theory-hardness}

Since a larger budget can make instances easier --- span-$0$ feasibility is NP-complete in general for $k\ge2$, yet polynomial on matching snapshots by Proposition~\ref{prop:span0} --- hardness of
\oneMinTCover{} for $k\ge2$ does not follow from the case $k=1$~\cite{ref_12}.

\begin{theorem}\label{thm:hardness}
\oneMinTCover{} is NP-hard for every fixed $k\ge1$.
\end{theorem}

\begin{proof}
For $k=1$ this is~\cite{ref_12}; let $k\ge2$. We reduce from the case $k=1$. Let $G$ be an instance with timestamps in $[T]$ and let $s\le n(T-1)$ be the span bound of the decision problem (larger bounds are satisfied by every instance); set $B=n(T-1)+1$. Build $G'$ by adding, for every $v\in V$, blocker vertices $c_{v,1},\dots,c_{v,k-1}$, each joined to $v$ by $k+1$ parallel temporal edges whose timestamps are pairwise more than $B$ apart, and more than $B$ away from $[T]$ and from the timestamps of every other blocker. All timestamps are distinct, so $G'$ is an \oneMinTCover{} instance of size polynomial in $n$, $T$ and $k$.

If $G$ admits a timeline of span at most $s$ with one interval per vertex, extend it to $G'$: each blocker covers $k$ of its $k+1$ edges by intervals of span $0$, and $v$ covers the remaining one by a further interval of span $0$, using $1+(k-1)=k$ intervals at $v$ and adding no span. Conversely, let $\mathcal{T}_k$ cover $G'$ with span at most $s<B$. No interval of $\mathcal{T}_k$ contains two timestamps more than $B$ apart, so each blocker, having at most $k$ intervals, covers at most $k$ of its $k+1$ pairwise $B$-separated edges; the remaining edge is covered by an interval of $v$ containing that blocker's timestamp, an interval which therefore contains no timestamp of $[T]$ and none of any other blocker. These $k-1$ intervals of $v$ are distinct, so at most one
interval of $v$ meets $[T]$, and restricting $\mathcal{T}_k$ to $[T]$ yields a timeline with one interval per vertex covering $G$ of span at most $s$.
\end{proof}

\subsection{Resource augmentation for the general problem}
\label{sec:theory-augment}

By Proposition~\ref{prop:barrier} no factor against $\OPT_k$ is attainable for general instances when $k\ge2$; the standard remedy is to compare against the optimum under a stricter budget. Write $\OPT_\kappa(G)$ for the minimum sum span over timelines with at most $\kappa$ intervals per vertex, and let $\mathrm{LP}(\kappa)$ be
\[
\min \sum_{v\in V}z_v
\quad\text{s.t.}\quad
x_u^t+x_v^t\ge1\ \ \forall\,(u,v,t)\in E,\quad
z_v\ge\textstyle\sum_{t\in[T]}x_v^t-\kappa,\quad
x_v^t\in[0,1],\ z_v\ge0,
\]
where $x_v^t$ indicates activity of $v$ at $t$ and $z_v$ its excess activity above the budget.

\begin{restatable}{theorem}{augment}\label{thm:augment}
For every $k\ge2$ and every $\theta\in(0,\tfrac12]$ with $\lfloor\theta k\rfloor\ge1$, a deterministic polynomial-time algorithm returns a covering timeline with at most $k$ intervals per vertex of sum span at most $\tfrac1\theta(T-1)\cdot\OPT_{\lfloor\theta k\rfloor}(G)$; in particular, at most $2(T-1)\cdot\OPT_{\lfloor k/2\rfloor}(G)$.
\end{restatable}

The threshold cannot exceed $\tfrac12$, since coverage only guarantees
$\max(x_u^t,x_v^t)\ge\tfrac12$. Whether the budget gap can be narrowed
from $k$ versus $\lfloor k/2\rfloor$ to $k$ versus $k-1$ is open, as is
a factor independent of $T$ in Theorem~\ref{thm:flow}, which would
extend the open question of~\cite{ref_B} to all $k$.

\begin{remark}\label{rem:dp-lemma5}
We note in passing that inequality~(10) in the analysis of the
randomized $O(T\log n)$ algorithm of~\cite{ref_B} passes from
$\sum_v\bigl(L\sum_t x_v^t-1\bigr)$ to
$L\sum_v\bigl(\sum_t x_v^t-1\bigr)$, where $L=\log_4c$ is the number of
rounding rounds; as the formulation enforces $\sum_t x_v^t\ge1$, the two
sides differ by up to $n(L-1)$, so the expected excess activity should
read $L\cdot\OPT_f+(L-1)n$ and the stated multiplicative factor does not
follow when $\OPT_f=o(n)$. None of the results above use randomized
rounding: Theorem~\ref{thm:flow} is exact at the level of the relaxation
on matching snapshots, and Theorem~\ref{thm:augment} uses a
deterministic threshold whose per-vertex accounting is loss-free at
$z_v=0$.
\end{remark}

\section{Exact per-vertex reconstruction}
\label{sec:heuristic}

Both the flow algorithm of Section~\ref{sec:theory-k} and the learned method of Section~\ref{sec: ML} follow the assignment view of Section~\ref{preliminaries}: they first fix an assignment $a$ of every temporal edge to one of its endpoints, and then give each vertex $v$ a cheapest timeline covering its required set $S_v(a)$. The second step is solved exactly by Lemma~\ref{lemma:minspan}, and we state it as Algorithm~\ref{alg:reconstruct}.

\begin{algorithm}[htb]
\caption{Exact per-vertex reconstruction}\label{alg:reconstruct}
\textbf{Input:} a set $S \subseteq [T]$ of timestamps and a budget $k \ge 1$.\\
\textbf{Output:} a timeline of at most $k$ intervals covering $S$, of span $\mathrm{minSpan}(S,k)$.
\begin{algorithmic}[1]
\If{$S = \emptyset$}
    \State \Return the empty timeline
\EndIf
\State sort $S$ as $t_1 < t_2 < \dots < t_p$
\State $C \gets$ the indices $i \in [p-1]$ of the $\min(k,p)-1$ largest gaps $t_{i+1}-t_i$, ties broken in favour of smaller $i$
\State cut the sequence $t_1,\dots,t_p$ after each index in $C$, obtaining $|C|+1$ blocks of consecutive elements
\State \Return the intervals $[\min B, \max B]$, one for each block $B$
\end{algorithmic}
\end{algorithm}

\begin{lemma}\label{lem:reconstruct}
Algorithm~\ref{alg:reconstruct} returns an activity timeline of at most $k$ intervals that covers $S$ and has span $\mathrm{minSpan}(S,k)$, in $O(p\log p)$ time, where $p=|S|$. Consequently, for every assignment $a$, applying it to each $S_v(a)$ yields a covering timeline of $G$ with at most $k$ intervals per vertex and sum span $\sum_{v}\mathrm{minSpan}(S_v(a),k)$, in $O(m + \sum_v |S_v(a)|\log|S_v(a)|)$ time.
\end{lemma}

\begin{proof}
For $S=\emptyset$ the empty timeline is optimal by definition. Otherwise the blocks are separated by gaps of length at least one, so the returned intervals are pairwise disjoint and ordered as required by the definition of activity timeline; there are $|C|+1=\min(k,p)\le k$ of them, every timestamp of $S$ lies in its block's interval, and the total span is $(t_p-t_1)-\sum_{i\in C}(t_{i+1}-t_i)$, which equals $\mathrm{minSpan}(S,k)$ by Lemma~\ref{lemma:minspan}. Sorting dominates the running time. The second statement follows from Observation~\ref{obs:assignment-feasible}: the union of timelines covering the sets $S_v(a)$ covers $G$, whatever the assignment $a$; computing the sets $S_v(a)$ takes $O(m)$ time.
\end{proof}

In particular, the feasibility of the solutions returned by our algorithms does not depend on how the assignment is chosen: every temporal edge is covered, and no vertex uses more than $k$ intervals. Our implementation computes the same optimum with the standard $O(p^2k)$ segmentation recurrence instead of the gap selection of Algorithm~\ref{alg:reconstruct}; both return a timeline of span $\mathrm{minSpan}(S_v(a),k)$, so the choice affects the running time only.

The conference version of this paper~\cite{mateiu2025time} also proposed a local-search step that recomputes the timeline of one vertex at a time on top of the heuristics of~\cite{ref_9}; we do not evaluate it here. We only note that, when the per-vertex step is solved exactly, such a recomputation never increases the span: once a vertex $v$ is removed, the timestamps of the temporal edges that become uncovered are all contained in the old timeline of $v$, which is therefore one of the candidate solutions.

\section{A Learning Based Approach}
\label{sec: ML}

\subsection{Preliminaries for the Learning Based Approach}

Below describe the necessary notations and definitions used in the presentation of the learning based approach. 

\begin{definition}
    An activity mask for a vertex $v$ is a binary vector $x_v \in { \{0,1\}}^T$ with $x_v^t = 1$ \textbf{iff} $v$ is active at timestamp $t$. 
\end{definition}

In our solution, we adopt the convention that $x_v^0 = x_v^{T+1} = 0$, padding the timeline with inactive borders at both ends. 

The transition count of a mask $x_v$ is defined as:

\begin{equation*}
    \tau(x_v)\;=\;\sum_{t=0}^{T}\big|x_v^{t+1}-x_v^{t}\big| ,
\end{equation*}

and is the number of positions at which the activity state changes between consecutive timestamps.

\begin{lemma}\label{lem:mask}
Let $v \in V$, let $\mathcal{T}_k(v)$ be an activity timeline of $v$ with $j$ intervals, let $A_v$ be its set of active timestamps and let $x_v$ be the corresponding activity mask. Then $\tau(x_v) = 2r(A_v) \le 2j \le 2k$. Conversely, every mask $x_v$ with $\tau(x_v) \le 2k$ is the activity mask of some timeline of $v$ with at most $k$ intervals, and the cheapest such timeline has span $|A_v| - \min(k,|A_v|)$.
\end{lemma}
\begin{proof}
With the padding $x_v^0 = x_v^{T+1} = 0$, $\tau(x_v)$ is the left-hand side of Fact~1 in the proof of Theorem~\ref{thm:ilp}, so $\tau(x_v) = 2r(A_v)$, and $r(A_v) \le j \le k$ by Fact~2(a). For the converse, $\tau(x_v) \le 2k$ gives $r(A_v) \le k$, and Fact~2(b) yields a timeline with active set $A_v$ and exactly $j$ intervals for every $j$ with $r(A_v) \le j \le \min(k,|A_v|)$; by Fact~2(a) its span is $|A_v| - j$, which is smallest for $j=\min(k,|A_v|)$.
\end{proof}
The correspondence between timelines and masks is not a bijection. As observed in Section~\ref{sec:ILP}, the timelines $\{[t,t+1]\}$ and $\{[t,t],[t+1,t+1]\}$ have the same mask but spans $1$ and $0$: a mask determines the set of active timestamps, but neither the intervals nor the span. For this reason our method never reads intervals off a mask; it reconstructs them exactly from the timestamps assigned to each vertex (Section~\ref{sec:heuristic}).

Lazarinetti et al. \cite{lazzarinetti2025dlmintc+} have already presented a deep learning approach for the Minimum Timeline Cover problem for the case where we have one interval $( k = 1)$. Our aim is to extend their solution for the general problem where $k$ can be any constant number. Their solution is organized into three functional parts: a learned encoder, which converts the temporal graph into a collection of vectors composed of vertex and timestamp $(v,t)$ pairs, a learned decoder, which reads the vectors and predicts the two endpoints for each vertex's activity interval and a deterministic repair step, backed by an algorithm which adjusts the intervals in order to make sure the result is guaranteed to cover all edges.

The encoder part is split in two stages. First, each vertex $v$ at every timestamp $t$ is described by its degree in the graph snapshot at time $t$. A graph neural network (GraphSAGE \cite{DBLP:journals/corr/HamiltonYL17}) then enriches this feature ($deg(v,t)$) by aggregating information from $v$'s neighbors s.t. the resulting vector reflects not only $v$'s degree but also its relevance in the graph at time $t$. This yields, for each vertex $v$ a sequence of per-timestamp embeddings $[h_{v}^{(1,K)}, h_{v}^{(2,K)}, \dots, h_{v}^{(T,k)}]$, where $T = |\mathcal{T}|$ is the number of timestamps in the timeline and $K$ is the final GraphSAGE layer. 

Second, sinusoidal positional encodings are added to preserve the ordering of timestamps and the sequence is passed through a Transformer encoder, which processes the sequence of a vertex's vectors across the timeline, producing a final embedding of the same length which shows how $v$'s connectivity changes across timestamps. The output vector at position $t$ is a weighted combination of information from all the other timestamps with the weights (attention) learned: $P_v = [P_v^{(t)} : t \in \mathcal{T}]\forall v_i \in V$.  

These embeddings are the input for the decoder part, which is represented by a Pointer Network \cite{vinyals2015pointer_arxiv}. The main objective of the Pointer Network, in the context of the \textit{MinTCover} problem, is to choose for each node in the temporal graph the most efficient activity interval, one that captures all relevant interactions while minimizing its span. More exactly, for each vertex $v$, the PN computes a probability distribution over the timestamps in $T$ and selects the two highest scoring positions as the start point $s_v$ and the end point $e_v$ of the activity interval of $v$. The intervals predicted by the Pointer Network are not guaranteed to form a valid cover, so Lazarinetti et al. introduces a deterministic post-processing step, the minimal timeline completion algorithm. This final step aims to have the activity intervals cover all temporal edges in $E$ i.e., for every $(u, v, t) \in E, t \in I_u \lor t \in I_v$, by extending the predicted intervals (either from the start or from the end). Lazarinetti et al. prove that the resulting intervals are locally minimal with respect to the increase in total span; that is, no final interval can be reduced without uncovering at least one temporal edge. 

The architectural choices of \textit{DLMinTC+} are well-matched to the single-interval setting but rest on three implicit assumptions that fail when $k > 1$. Identifying these assumptions is essential both because it explains why a direct extension is unsatisfactory and because it motivates the design of our method.
The first assumption is structural: the output of each vertex can be encoded by exactly two scalars. The Pointer Network architecture commits to producing precisely two pointers — one for $s_v$, one for $e_v$ — and the loss function compares these two scalars to ground truth pairs $(\hat{s}_v, \hat{e}_v)$ via mean squared error. For $k > 1$, a vertex's timeline is no longer characterized by a fixed-dimensional output. Any direct generalization that retains the pointer-based output must commit to one of two architecturally costly options: predict a fixed-size sequence of $2k$ pointers and rely on auxiliary mechanisms to handle the variable effective number of intervals (e.g., via end-of-sequence tokens), or move to a fully autoregressive decoder that emits one pointer at a time. Both options, unfortunately, re-introduce the computational complexity that the pointer network used for $k=1$ aimed to avoid. The second assumption is regarding the loss. In the case for $k=1$,  the only nontrivial structural constraint on the output is $s_v \le e_v$, which Lazzarinetti et al. enforce with a soft penalty function $\text{ReLU}(s_v - e_v)^2$. For a bigger $k$, we must consider that the predicted timeline must consist of at most $k$ time-disjoint intervals, which in mask form translates to a bound on the number of activity transitions across the timeline. Encoding this constraint via penalties on raw pointer outputs is awkward — it requires sorting predicted intervals, computing pairwise gaps, and penalizing overlaps — and the resulting loss landscape is poorly behaved near the valid region. The third assumption is in the post-processing: that local extension of existing intervals is a sufficient operation. The minimal timeline completion algorithm of DLMinTC+ handles uncovered edges by extending the nearer of the two endpoints $s_v$ or $e_v$ to include the timestamp of the uncovered vertex. For $k>1$ this might not be the most optimal option. To extend the method to general $k$ we must replace each of the assumptions with a formulation that natively accommodates a variable number of disjoint intervals.

\subsection{Problem Principles}
\label{principles}

We frame our extension around three design principles, each responding to one of the assumptions identified above. 

- Firstly, regarding the structure of the activity timeline, in order to adapt it to the $k$ intervals, the activity must be bijection with a binary mask $x_v \in \{0,1\}^T$ whose number of active-deactive and deactive-active transitions is at most $2k$. The network emits $T$ real numbers per vertex, one per timestamp: $(p_v^1, p_v^2, \dots p_v^T)$.  Each number can be interpreted as the probability that vertex v should be active at time t. To convert these continuous probabilities into a discrete binary mask, we apply a threshold: values at or above $0.5$ become $1$ (active), values below become $0$ (inactive). The resulting binary vector is the mask and the intervals are recovered by reading off the maximal runs of $1$ values in it. This way, the mask is bounded by the input graph (specifically $T$) instead of being bounded by $k$ (which scales independently of the graph structure) which would have happened if we decided to go forward with the Pointer Network approach used for \textit{DLMinTC+}. The advantage here is that the transformer encoder already produces $T$ embeddings per vertex regardless of how the output head is designed, so a per-timestamp output head simply applies one additional linear projection at each of those $T$ positions. 

 - Second, we encode the constraints in differentiable form, but will not rely on the network alone to satisfy them. In order to solve the \textit{k-MinTimelineCover} problem we must satisfy two constraints: every temporal edge must have at least one active endpoint and every vertex must have at most $k$ intervals. Thus, we add penalty points to the loss that discourage invalid outputs - uncovered edges and vertices exceeding the interval budget $k$ - but we do not require the network's raw output to be feasible. Enforcing strict feasibility is pushed to the post-processing step. The network's role remains to supply a high quality but approximate output. 

 - Third, we use the post processing step to do more than fixing the coverage. For $k=1$, repair step has a single job: extend the nearer interval endpoint to include uncovered edges. For $k > 1$ the structure is richer and the set of natural repair operations is correspondingly larger. Concretely, consider for example a vertex $v$ with one predicted interval $[3,7]$ and an uncovered edge requiring $v$ to be active at timestamp $20$. If $k=1$, the only fix would be to extend the interval to $[3,20]$. For a bigger $k$, we can add a new interval $[20,20]$ with $0$ cost. The post processing step from \textit{DLMinTC+} has no way to express operations other than extending a single interval. Our post processing therefore decides, for each vertex, how to group the needed activity timestaps into $k$ intervals - including creating new ones.

\section{Outlie of k-DLMinTC+}

Combining the three design principles yields the foundation of the \textbf{k-DLMinTC+} pipeline:
 \begin{enumerate}
     \item Encoding: GraphSAGE is applied to the temporal graph at every moment in time $t \in 1 \dots T$, producing static embeddings $h_v^{(t,K)}$ for each $(v,t)$ pair. These will be stacked into per vertex sequences and then processed by a Transformer with sinusoidal positional encoding, producing the temporally contextualized embeddings $P_v^t$. 
     \item Prediction: A small feedforward head maps each $P_v^{(t)}$ to a scalar $p_v^t \in [0,1]$ via sigmoid activation. The collection $\{p_v^t\}_{v,t}$ defines a soft binary mask over the temporal graph.
     \item Post-processing. The soft mask is converted into a feasible solution in two steps. First, every temporal edge $(u,v,t)$ is assigned to exactly one of its endpoints - the vertex responsible for covering it - which determines for each vertex $v$ the set of timestamps $S_v$ it must be active at. Second, independently for each vertex, an $\mathcal{O}(|S_v|^2 \cdot k)$ dynamic program computes the minimum span set of at most $k$ intervals covering $S_v$.
 \end{enumerate}

 \subsection{Encoding Algorithm}
    
    In this subsection we formally present the encoding part. The encoder maps a temporal graph $G = (V, E)$ and the intervals $k$ to one contextualized embedding per $(v, t)$ pair. It reuses the two-stage structure of DLMinTC+ from Lazarinetti et al. - a per-snapshot GraphSAGE module, and a temporal Transformer, with one addition which happens first: because a single trained model must serve all activity intervals $k$, the interval budget has to enter the network somewhere. Thus, we start by giving each value of $k$ its owned learned vector $e_k$ and add it into the input features.
    The first two components  produce static, per-timestamp representations; the Transformer makes them aware of the whole temporal context producing a final, aggregated output. We describe the construction of each component in turn as follows: 
    \subsubsection{Conditioning on k} 
    In order to accommodate the fact that now $k>1$, the interval budget must be supplied to the network as input. We do not wish to feed it as a raw integer $k$, thus we map $k$ to a learned vector. Let $k_{max}$ be the value of the maximum number of intervals a vertex can be active in. Let $\mathbb{R}^{(k_{\max}+1) \times d_k}$ be the space of real matrices with $k_{\max}+1$ rows and $d_k$ columns and let $E_k$ be such a matrix whose entries are trainable parameters. We call $E_k$ an \emph{embedding table}: it has one row for each admissible budget $i \in \{0,1,\dots,k_{\max}\}$, and $d_k$ is the embedding dimension, a hyperparameter which we set during the implementation. The budget $k$ is then represented by the $k$-th row of the table, i.e. $e_k = \mathbf{1}_k^{\top},$ where $\mathbf{1}_k \in \{0,1\}^{k_{\max}+1}$ is the one-hot indicator of $k$. Writing the lookup as this linear projection makes explicit that the operation is differentiable, so the entries of $E_k$ are learnt jointly with the rest of the network. The resulting vector $e_k$ is concatenated to every $(v,t)$ input feature, so that the entire network is conditioned on the interval budget. It is important to note that the rows of $E_k$ are not hand-designed and do not have predefined semantics: they are randomly initialized, trainable parameters whose values are determined solely by minimizing the training objective. Functionally, $e_k$ acts as a mode selector that lets a single shared encoder adapt its per-$(v,t)$ activity predictions to the budget it is currently solving for, producing more selective covers, concentrated into few contiguous intervals, for small $k$, and more fragmented covers for large $k$. We prefer this learned embedding to a raw scalar input because the latter would force the network to assume a fixed functional dependence on the magnitude of $k$, whereas an embedding assigns each budget its own free vector and provides a higher-capacity conditioning signal, following standard practice for conditioning neural networks on discrete inputs.

    \subsubsection{Per-snapshot structural embedding.}
    The temporal graph is viewed as a sequence of snapshots ${G_t} , t=1,\dots T$ with $G_t = (V, E_t)$ and $E_t = \{(u,v) : (u,v,t) \in E\}$. The initial feature of vertex $v$ in snapshot $t$ is its snapshot degree $deg(v, t)$. We treat this number as a length one vector and \emph{concatenate} it with the interval budget vector $e_k$ - i.e. we stack the two into a single vector of length $1 + d_k$ which serves as the input feature of $v$ at timestamp $t$. We apply $k$ SAGEConv \cite{DBLP:journals/corr/HamiltonYL17} layers; One GraphSAGE layer updates the representation of $v$ in two operations: it summarizes the representation of $v$'s neighbors into a single vector using a mean aggregator, and it combines it with $v$'s own representation through a learned linear transformation and a non linear activation. Concretely, the layer-$\kappa$ update for vertex $v$ in timestamp $t$ is:

\begin{equation}
    h_v^{(t,\kappa)} = \mathrm{ReLU}\left( W^{(\kappa)} \cdot \frac{1}{|N_t(v)| + 1} \sum_{u \in N_t(v) \cup \{v\}} h_u^{(t,\kappa-1)} \right)
    \label{eq:graphsage-update}
\end{equation}

    where $N_t(v)$ is the neighborhood of $v$ in timestamp $t$, $W^{(\kappa)}$ is the trainable weight matrix of layer $\kappa$, $\mathrm{ReLU}(x) = \max(0,x)$ is the activation function and $h_v^{(t,0)}$ is the input feature. The output $h_v^{(t,K)}$ summarizes v's local connectivity in snapshot t. We include each vertex's representation in the average - hence $\cup \{v\}$ in the sum - s.t each $v$'s current representation contributes to its next one. 
    
    To avoid an explicit loop over timestamps, all snapshots are processed in one batched pass over a disjoint-union "supergraph" on vertex set $V \times [T]$, in which messages pass only within a snapshot; this yields a cost of $\mathcal{O}(|E| · K)$ independent of T.

\subsubsection{}{Temporal Transformer.} The per-timestamp embeddings are assembled into one sequence per vertex,

\begin{equation*}
    H_v = \left[\, h_v^{(1,K)},\ h_v^{(2,K)},\ \dots,\ h_v^{(T,K)} \, \right]
\end{equation*}

where $T = |\mathcal{T}|$. Because self-attention is invariant to input order, sinusoidal positional encodings (Vaswani et al. \cite{vaswani2023attentionneed}), computed by a fixed formula so as to admit arbitrary T, are added componentwise to give $\hat{H}_v$. The sequence is then passed through $L$ Transformer encoder layers, each having a multi-head self-attention sublayer followed by a position-wise feedforward sublayer, with residual connections and layer normalization. Attention is applied independently within each vertex's temporal sequence — no attention runs across vertices at this stage — which keeps the cost at $O(T^2)$ per vertex rather than $O((|V|T)^2)$. The output is the contextualized sequence

\begin{equation*}
   P_v = [ P_v^{(t)},  t \in \mathcal{T}]
\end{equation*}
    
computed for every $v \in V$, in which each $P_v^{(t)}$ is informed by v's activity across the whole timeline rather than by timestamp t alone.

Restricting attention to within-vertex sequences means cross-vertex temporal dependencies are mediated only by the GraphSAGE component. A vertex's optimal timeline does depend on its neighbors' choices, so this is a deliberate simplification, accepted for two reasons: full spatiotemporal attention would be $O((|V|T)^2)$ and prohibitive at the instance sizes of interest, and the post-processing stage coordinates across vertices explicitly, by assigning each edge's coverage responsibility based on the actual predicted activity rather than on the embeddings alone.

\subsection{Prediction}

The encoder gives us, for every vertex and every timestamp a vector that summarizes the vertex's local connectivity and how its activity evolves over the timeline. What remains is to turn these vectors into an actual decision: when the vertex is active in the timeline. This section describes how these decisions are made and how the network is trained to make the decisions optimal.

Comparing with the case where $k = 1$, rather than having the prediction output the endpoints of each interval directly - we have our network predict, independently for every timestamp, wether the vertex is active then. The reasons were given in Subsection~\ref{principles}: a per-timestamp prediction consists in one probability per timestamp, having $T$ values per vertex - an output which is the same for every $k$ - so a single architecture would serve any interval number. 

Another important aspect is that we do not train the network to produce a valid cover outright. A valid cover obtained right from the training phase would be difficult to obtain, but also unnecessary, since the prediction is followed by a post-processing step which aims to repair whatever is left uncovered by the network. The network only needs to get close, and the loss is designed accordingly. The prediction head itself is small. A two-layer feedforward network is applied to each embedding $P_v^{(t)}$, and its output is passed through a sigmoid function producing 
\begin{equation}
p_v^t = \sigma\big(\mathrm{MLP}(P_v^{(t)})\big) \in (0,1),
\end{equation}

which we read as the probability that vertex $v$ is active at timestamp $t$. The same head is applied at every $(v,t)$ pair, so one forward pass yields the full matrix $p \in (0,1)^{|V| \times T}$ Thresholding at $0.5$ converts $p$ into a binary mask whose maximal runs of ones are the predicted intervals — but this mask is a starting point, not a solution, and next section presents how it becomes one. 

The network is trained by minimizing a weighted sum of four differentiable terms: one supervised term that teaches it to imitate known good solutions, and three that encode the structure of the problem itself. For the supervised term, each training graph is paired with a reference solution — the exact optimum computed by the ILP when it terminates within a time budget, and the best of the available heuristics otherwise - recorded as a binary mask $\hat{x} \in \{0,1\}^{|V| \times T}$ marking which vertices get activated at which timestamps. The imitation term is the binary cross-entropy between the network's probabilities and this mask:
\begin{equation}
    \mathcal{L}{\mathrm{imit}} = -\frac{1}{|V|,T}\sum{v,t}\Big[\hat{x}_v^t\log p_v^t + (1-\hat{x}_v^t)\log(1-p_v^t)\Big]
\end{equation}

Since $\hat{x}$ can be either 0 or 1, exactly one of the two values can survive for each pair: the loss charges $-log(p_v^t)$ where the reference is active and $-log(1-p_v^t)$ where it is not. Minimizing $\mathcal{L}{\mathrm{imit}}$ is exactly maximizing the likelihood the network assigns to the reference mask. Reading $p_u^t$ and $p_v^t$ as independent probabilities, the probability that the edge $(u,v,t)$ is uncovered is $(1-p_u^t)(1-p_v^t)$, so the probability that it is covered is $p_u^t + p_v^t -p_v^t p_u^t$ and we penalize its negative logarithm:

\begin{equation}
    \mathcal{L}{\mathrm{cov}} = -\sum_{(u,v,t)\in E}\log\big(p_u^t + p_v^t - p_u^t p_v^t + \varepsilon)
\end{equation}

including the constant $\varepsilon$ for numerical stability. The sparsity term pulls in the opposite direction, penalizing activity that coverage does not require, so that the network cannot satisfy coverage trivially by activating everything:

\begin{equation}
    \mathcal{L}{\mathrm{spar}} = \frac{1}{|V|,T}\sum{v,t} p_v^t.
\end{equation}

The tension between these two terms is deliberate: coverage forces enough activity to cover the edges, sparsity keeps it no larger than necessary, and their balance is what a low-span cover looks like at the level of soft probabilities.

The last term enforces the interval budget. A binary mask corresponds to at most $k$ intervals exactly when it switches between active and inactive at most $2k$ times. The switch count of the binary mask is $\small\sum_t |x_v^{t+1} - x_v^t|$, which is not differentiable. Thus, we apply the same expression to the soft probabilities, where it coincides with the discrete count whenever $p$ takes values in $\{0,1\}$, and penalize only the excess beyond the budget:
\begin{equation}
    \mathcal{L}{\mathrm{trans}} = \sum_v \mathrm{ReLU}\Big(\sum{t}\big|p_v^{t+1} - p_v^t\big| - 2k\Big)
\end{equation}
A vertex within its budget contributes nothing to this term, so it interferes with the rest of the loss only where it must.
The total loss is described as the following equation:

\begin{equation}
\mathcal{L} = \lambda_{\mathrm{imit}}\mathcal{L}{\mathrm{imit}} + \lambda{\mathrm{cov}}\mathcal{L}{\mathrm{cov}} + \lambda{\mathrm{spar}}\mathcal{L}{\mathrm{spar}} + \lambda{\mathrm{trans}}\mathcal{L}_{\mathrm{trans}}
\end{equation}

with every term differentiable, so the whole objective is trained by gradient descent. The weights $\lambda$ together with a schedule that phases the budget term in gradually - letting the network first learn to cover edges before it learns to economize on transitions - are given in Section~\ref{sec:experiments}. None of the four terms guarantees a feasible output, and none needs to: their job is to produce a soft mask close enough to a good cover that the post-processing has little left to repair.

\subsection{Post-processing}
\label{sec:postproc}

The network's output $p \in (0,1)^{|V| \times T}$ is not yet a solution. Thresholding it may leave edges uncovered, and a vertex's mask may exceed its transition budget. Post-processing turns $p$ into a feasible cover in two steps: it first fixes an edge assignment which we will name $a$, and then solves each vertex's covering problem exactly. 

Once $a$ is fixed, the vertices decouple: each vertex $v$ must cover its required set $S_v(a)$. Given $S_v(a)$ and the budget $k$, we need at most $k$ disjoint intervals covering $S_v(a)$ with a minimum total span $minSpan(S_v(a), k)$. This is precisely the problem solved in algorithm \ref{alg:alg1} in $\mathcal{O}(n^2k)$ time by the dynamic program of Section \ref{sec:heuristic}. We apply it unchanged, with $U = S_v(a)$. This leaves us on deciding how to choose the assignment $a$. The objective suggests an immediate answer: route each edge to whichever endpoint can absorb it more cheaply. Concretely, maintain each vertex's assigned set as the edges are processed and send $(u,v,t)$ to the endpoint minimizing the marginal cost $\mathrm{minSpan}(S,k)$. 

This greedy solution is nonetheless not the optimal choice. The marginal cost of adding a timestamp to a vertex is zero whenever that vertex still has a spare interval to spend on it — and early in the processing order this holds for almost every vertex, since almost every vertex is still nearly empty. In this regime the greedy rule has no signal: both endpoints cost zero, the decision falls to an arbitrary tiebreak, and because costs are evaluated against the running assignment, these arbitrary early commitments then condition every subsequent evaluation. The rule is both order-dependent and, exactly where it decides the most, uninformed. Its characteristic failure is to \emph{scatter}: responsibility spreads thinly across many vertices, each accumulating isolated timestamps, and the dynamic program — exact though it is — can only wrap scattered points in wide or numerous intervals.
Deciding which vertices should be present in activity intervals is a global question about the instance, and it is not one that any local marginal cost can answer. Through the network results we can see which activity patters of the near-optimal solutions are and use them as reference. Routing by the prediction results therefore sends each edge toward a vertex the inferenced solution already uses. We adopt the following rule:
\begin{equation}
    a(u,v,t)=\arg\max\big(p_u^t,p_v^t),
\end{equation}
letting the learned probability decide responsibility and letting span enter only afterwards, through the dynamic program. This way, the design separates two decisions that the greedy approach treated as only one. \emph{Which} vertex covers an edge, which becomes a global question answered by the network, and \emph{How} a vertex's assigned timestamps are packed into intervals, which is treated by the DP algorithm. As we first went with the first approach in our implementation, we show a comparison of the two in the experiments (\ref{sec:experiments}) section which validates the major improvement that the probability-based rule produced.

\section{Experiments} 
\label{sec:experiments}



We evaluate the learned pipeline in three settings.
Section~\ref{sec:exp-setup} describes the data generation, the training protocol, the baselines, and the exact reference for the optimal solution. Section~\ref{sec:exp-social} evaluates on synthetic temporal graphs with realistic social structure at sizes where every instance can be solved to proven optimality, and isolates where the improvement comes from. Section~\ref{sec:exp-scaling} scales the evaluation first to instances up to $1.5$ times larger than the largest training instances, where the exact solver is two orders of magnitude slower than the learned pipeline, and then to instances with up to $5{,}000$ vertices, $25$ times larger, where it no longer certifies optimality. Section~\ref{sec:exp-highschool} applies the pipeline, without retraining, to a real-world contact network. 
\subsection{Experimental setup}
\label{sec:exp-setup}
A detailed view of the experiments setup can be found in Appendix~\ref{app:experiments-setup}


\paragraph*{Metrics.}
Every method considered returns a \emph{feasible} cover on every instance
--- in our pipeline each temporal edge is assigned to one of its endpoints
and the per-vertex dynamic program covers all assigned timestamps by
construction --- so coverage equals $1.0$ throughout and is not
discriminative. The reported quantity is the total span
$S(\mathcal{T}) = \sum_{v \in V} \sum_{I \in \mathcal{T}_v}
(\mathrm{end}(I) - \mathrm{start}(I))$, i.e.\ the summed \emph{duration}
of all activity intervals, under which a degenerate single-step interval
is free, and, per instance, the approximation ratio
$S_{\mathrm{method}} / S_{\mathrm{OPT}}$ against the exact optimum.

\paragraph*{Optimum and baselines.}
The optimum is computed with a binary program over activity variables
$x_{v,t}$ and transition variables $y_{v,t}$. So that the reference
optimises \emph{exactly} the reported metric, the objective is the total
duration
\[
\min \;
\sum_{v,t} x_{v,t}
\;-\; \tfrac{1}{2}\sum_{v} \Bigl(\sum_{t} y_{v,t} + x_{v,0} + x_{v,T-1}\Bigr),
\]
where $y_{v,t} = |x_{v,t+1} - x_{v,t}|$ is enforced exactly by
linearisation from both sides, the subtracted term counts one unit per
maximal activity run, and the interval budget is
$\sum_t y_{v,t} + x_{v,0} + x_{v,T-1} \le 2k$, which caps the number of
intervals per vertex at $k$ including runs touching the timeline
boundaries; a vertex may also remain entirely inactive. Each edge
constraint $x_{u,t} + x_{v,t} \ge 1$ enforces coverage. Gurobi records
whether optimality was \emph{proven}; approximation ratios are reported
only against proven optima, and where a proof is unavailable we
additionally report ratios against the solver's proven lower bound, which
upper-bound the true ratio.

We compare against the exact solver and against the three heuristics of
Rozenshtein et al.~\cite{ref_9bis} --- \emph{k-Inner}, \emph{k-Budget},
and \emph{k-Baseline} --- the strongest published non-learning baselines
for this problem family. Of our pipeline we evaluate three variants that
share the identical trained network and forward pass and differ only in
the edge-assignment rule of Section~\ref{sec:postproc}: the proposed
model-driven assignment (\emph{model}), the span-cost assignment that
consults the network only to break ties (\emph{greedy}), and their
combination (\emph{hybrid}).

\subsection{Results on synthetic social networks}
\label{sec:exp-social}

On the small test split the exact solver proves optimality on all
$200$ instances (mean solve time $1.4$\,s, at most $16.2$\,s), and every optimum is positive, so every ratio in
Table~\ref{tab:social-main} is measured against the true optimum.

\begin{table}[h!]
\centering
\caption{Small social test split ($200$ instances, $n \in \{30,50\}$,
$k \in \{2,\dots,10\}$; all optima proven; coverage $= 1.0$ for every
method). Mean ratio: mean of the per-instance empirical ratios, with $95\%$ bootstrap confidence interval. Ratio of totals: sum of spans over sum of optima.}
\label{tab:social-main}
\begin{tabular}{lrrrrr}
\hline
Method & Avg.\ span & Mean ratio [95\% CI] & Median & Worst & Ratio of totals \\
\hline
ILP (optimal)   & 1{,}341.0 & $1.000$ & $1.000$ & --- & $1.000$ \\
\textbf{ML (model)} & \textbf{1{,}519.0} & $\mathbf{1.134}$ $[1.125, 1.143]$ & $\mathbf{1.138}$ & $\mathbf{1.56}$ & $\mathbf{1.133}$ \\
ML (hybrid)  & 1{,}946.8 & $1.793$ $[1.639, 1.967]$ & $1.256$ & $7.97$ & $1.452$ \\
k-Inner         & 2{,}912.3 & $2.575$ $[2.427, 2.740]$ & $2.199$ & $8.76$ & $2.172$ \\
ML (greedy)  & 2{,}697.9 & $2.596$ $[2.373, 2.832]$ & $1.939$ & $10.69$ & $2.012$ \\
k-Baseline        & 3{,}675.7 & $3.635$ $[3.328, 3.965]$ & $2.847$ & $14.94$ & $2.741$ \\
k-Budget        & 7{,}241.2 & $7.282$ $[6.644, 7.977]$ & $5.561$ & $29.35$ & $5.400$ \\
\hline
\end{tabular}
\end{table}

\begin{figure}[h!]
\centering
\includegraphics[width=0.8\textwidth]{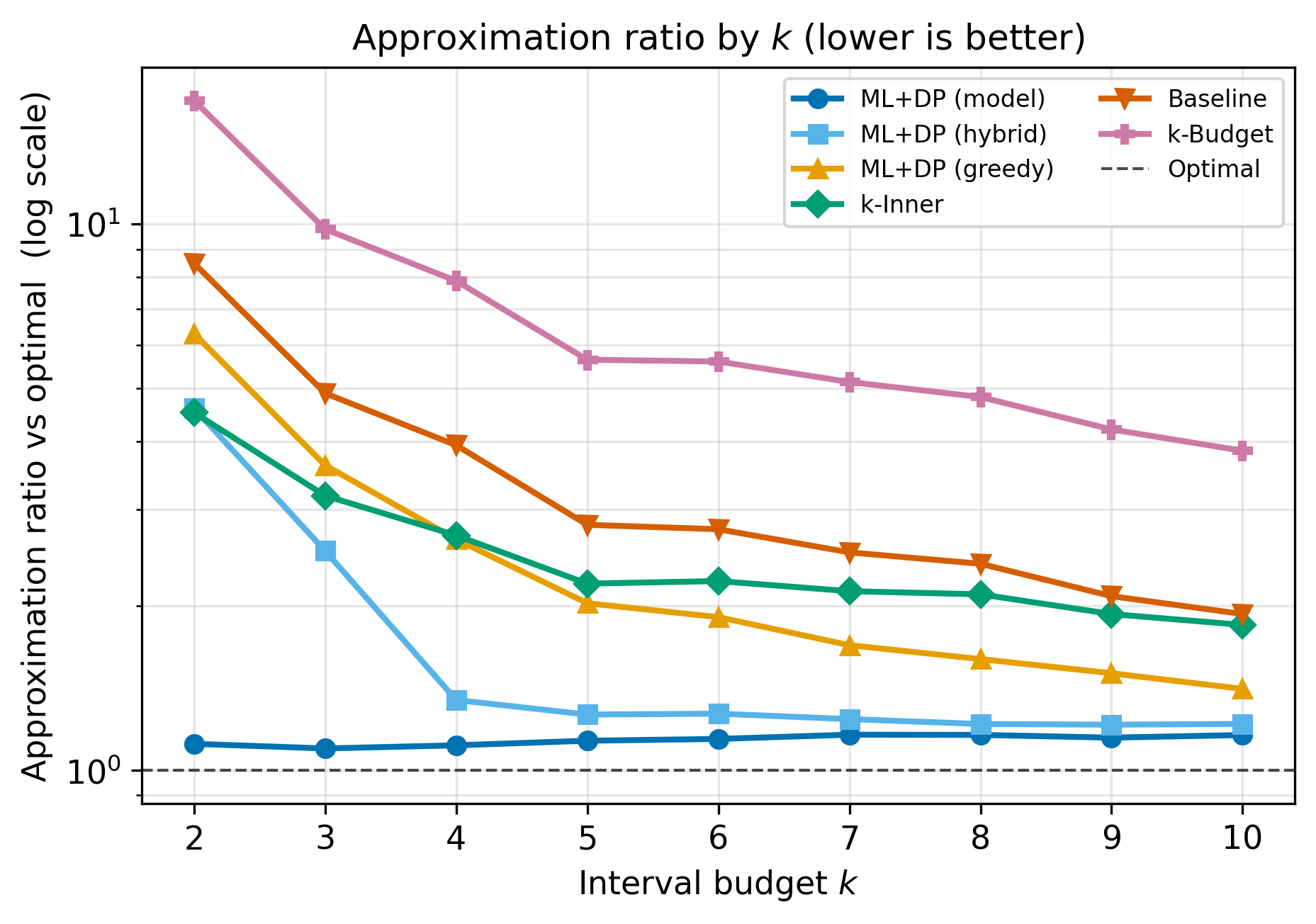}
\caption{Mean approximation ratio versus the proven optimum by interval
budget $k$ on the social test split (log scale). The proposed model-driven
pipeline is uniformly close to the optimum; all other methods drift upward
as $k$ shrinks.}
\label{fig:social-ratio-by-k}
\end{figure}

With the model assignment, the learned pipeline has a mean empirical ratio of $1.134$ (95\% confidence interval $[1.125, 1.143]$) on this split, with a worst case of $1.56$ over the $200$ instances. Over the three trained networks the mean ratio is $1.148 \pm 0.014$ (between $1.134$ and $1.161$). Compared instance by instance with k-Inner, the strongest heuristic, it returns a smaller span on all $200$ instances, with a mean reduction of $50.7\%$ ($[48.8\%, 52.7\%]$) and a reduction of the total span of $47.8\%$; the same holds on all $200$ instances for each of the three networks. It is the best non-exact method at every value of $k$
(Figure~\ref{fig:social-ratio-by-k}): its mean ratio per value of $k$ stays between $1.10$ and $1.16$, while the other methods degrade as $k$
shrinks and the problem becomes more tightly constrained (at $k = 2$:
$1.12$ for our method against $4.52$ for k-Inner). Its ratio distribution is concentrated just above $1$, with a short upper tail, whereas the worst cases of the heuristics are an order of magnitude above the optimum. One forward pass plus decoding takes $0.05$\,s per instance on average, against $1.4$\,s for the exact solver.

\begin{figure}[h!]
\centering
\includegraphics[width=0.8\textwidth]{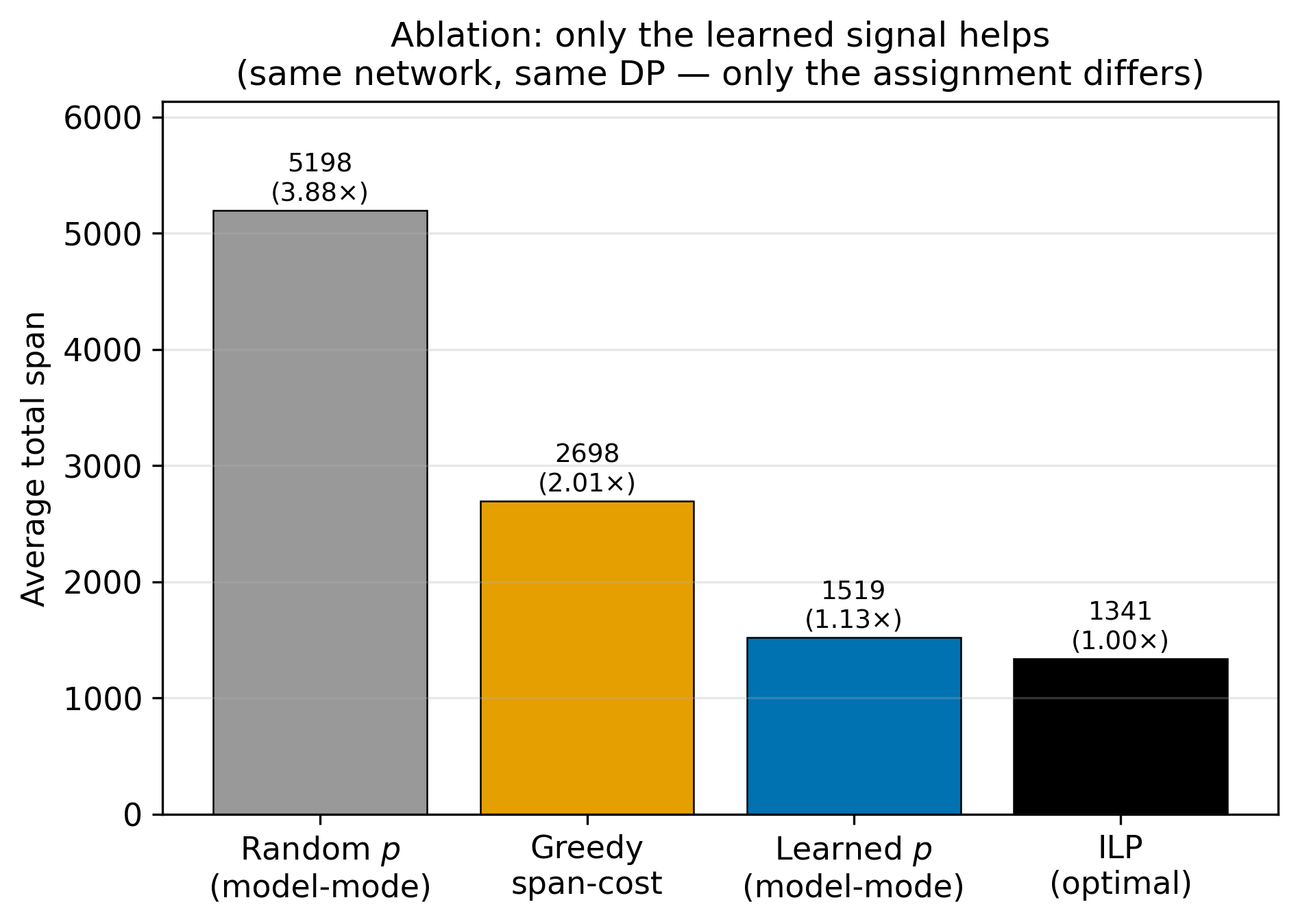}
\caption{Ablation on the social test split (ratios of average spans):
random probabilities under model-mode decoding, greedy span-cost
assignment, learned probabilities under model-mode decoding, and the
optimum. Only the combination of learned signal and model-driven decoding
approaches the optimum.}
\label{fig:social-ablation}
\end{figure}

\paragraph*{Where the improvement comes from.}
Because the three variants share the same trained network,
forward pass and reconstruction, differences among them are
due to the assignment step alone. Two comparisons isolate the
gain (Figure~\ref{fig:social-ablation}). First, replacing the
model assignment by the greedy span-cost assignment, leaving
everything else unchanged, raises the mean ratio from $1.134$
to $2.596$, slightly above that of k-Inner; the model assignment is better on all $200$ instances, with a mean span reduction of $44.4\%$ ($[41.4\%, 47.5\%]$). A decoder that consults the network only to break ties discards essentially all learned information.
Second, to exclude the possibility that the model assignment is strong
by itself, we replaced the network's output by scores drawn independently and uniformly from $(0,1)$, ten times per instance with different seeds, and decoded each draw with the model assignment. The ratio of totals rises to $3.88$ (range over the ten draws $[3.87, 3.88]$) and the mean ratio to $5.03$ ($[5.00, 5.06]$), against $2.01$ and $2.60$ for the greedy assignment and $1.13$ and $1.13$ for the learned scores. The decoder is therefore not the source of the
improvement but the channel through which the learned predictions reach
the solution: a good network is of no use if the decoder ignores it, and the
decoder is of no use if the network carries no signal.

An earlier network with the same architecture was trained on $2{,}000$ instances with $n \in \{30, 50, 80\}$ and horizons of at least $225$, whose labels were exact solutions on only about $10\%$ of the instances and the best of the three heuristics on the others. Its mean empirical ratio is $1.220$ ($[1.208, 1.233]$) on the small split and $1.239$ ($[1.228, 1.251]$) on the large one, against $1.134$ and $1.141$ for the network used here, and it is also worse on every real data set of Sections~\ref{sec:exp-highschool} and~\ref{sec:exp-robust}, most markedly on the $1$-hour bins, where it returns a smaller span than k-Baseline on $8$ of the $49$ instances, against $41$. The two training sets differ at once in the source of the labels, in the sizes and in the horizons, so this comparison does not attribute the gain to any one of these factors.

\subsection{Scaling to larger instances}
\label{sec:exp-scaling}

\begin{figure}[h!]
\centering
\includegraphics[width=\textwidth]{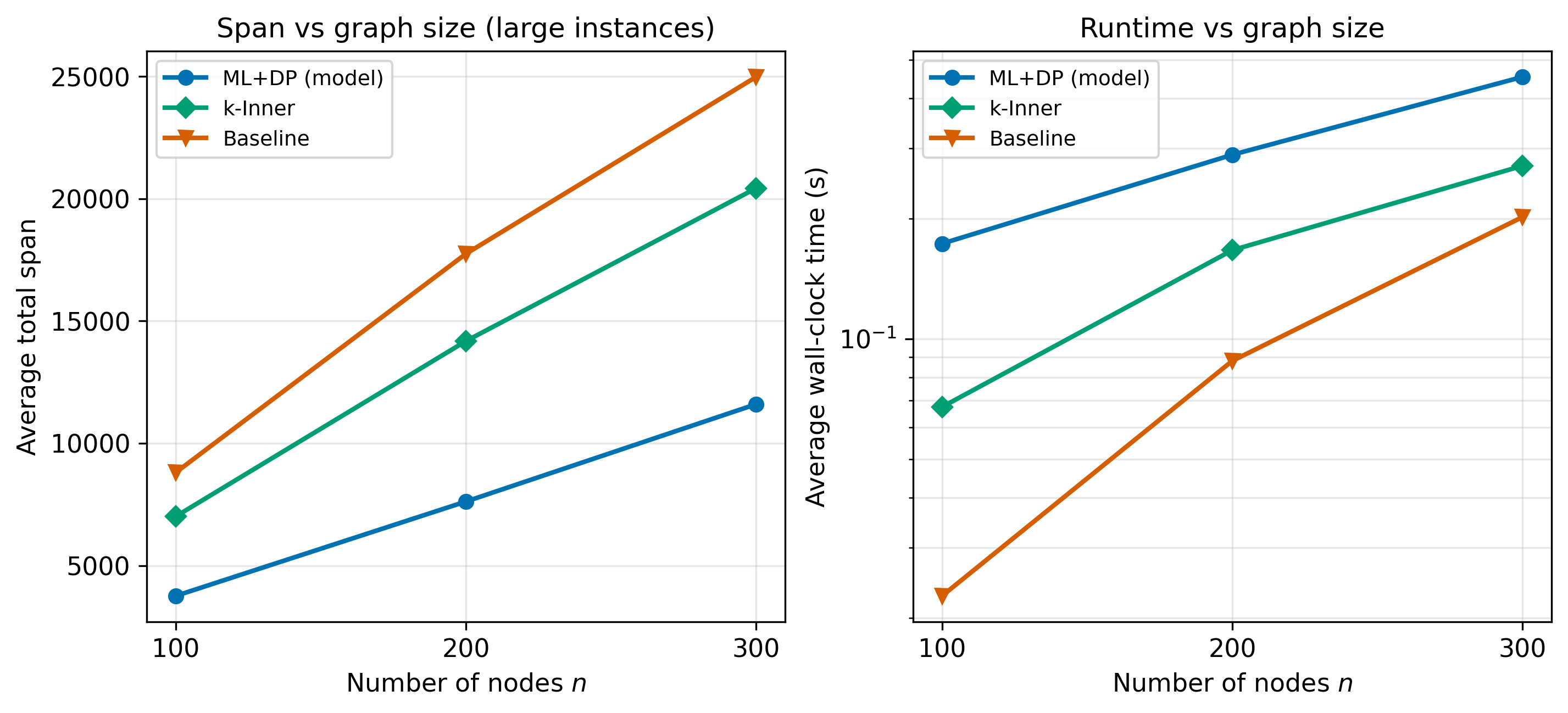}
\caption{Mean span (left) and mean wall-clock time (right, log scale)
versus graph size on the large social split. The learned pipeline
keeps a constant-factor advantage over the heuristics at sub-second
cost.}
\label{fig:scaling}
\end{figure}

The same trained network, \emph{without} retraining, is then
evaluated on the large split, whose instances have up to $1.5$ times more vertices than the largest training instances. Here the exact solver is no longer comfortable: with a
$120$\,s time limit it proves optimality on $182$ of the $200$ instances ($65$ of $66$ with $n=100$, $52$ of $56$ with $n=200$, and $65$ of $78$ with $n=300$), with a mean running time of $21.5$\,s per instance (median $6.4$\,s), whereas one forward pass plus decoding takes $0.31$\,s. Since the solver is stopped at $120$\,s, this ratio of running times is a lower bound on the speed difference for proving optimality. On each of the $18$ remaining instances the solution returned by the solver is within $0.08\%$ of its lower bound, so these instances are solved in practice and only the certificate is missing; with a limit of $30$\,min, $12$ of them are proven as well, so that $194$ optima are proven ($66$ of $66$, $55$ of $56$ and $73$ of $78$), and the remaining $6$ solutions are within $0.05\%$ of their lower bounds. Doubling the time limit matters, however: with a limit of $60$\,s, $26$ instances remain unproven, and on three of them the solver returns solutions $3.6$ to $4.6$ times larger than the optimum, worse than those of the model assignment.
Table~\ref{tab:scaling} and
Figure~\ref{fig:scaling} summarize the results. On the proven instances the mean empirical ratio of the model assignment is stable as $n$ grows ($1.132$, $1.147$ and $1.143$ for $n = 100, 200, 300$; $1.141$, $[1.134, 1.147]$, overall, over the $194$ proven instances, and $1.162 \pm 0.019$ over the three networks), and compared instance by instance with k-Inner it returns a smaller span on all $200$ instances, with a mean reduction of $47.7\%$ ($[45.6\%, 49.7\%]$), as does each of the three networks. To include the $6$ instances without a proven optimum, we also divide each span by the best lower bound found by the solver, which can only overestimate the true ratio: over all $200$ instances the mean of these ratios is $1.141$ ($[1.135, 1.148]$) for the model assignment. On this split, the gap to the optimum is therefore the same at training sizes and beyond them.

\begin{table}[h!]
\centering
\caption{Large social test split ($200$ instances,
$n \in \{100, 200, 300\}$, $k \in \{2,\dots,10\}$). Mean empirical ratios to the proven optimum by $n$ ($194$ proven instances, $182$ of them within $120$\,s and $12$ more within $30$\,min); mean ratio to the solver's lower bound (to the optimum, when proven) over all $200$ instances; average span over the $194$ proven instances; average time over all instances, for the $120$\,s run; average time over all instances.}
\label{tab:scaling}
\small
\begin{tabular}{lrrrrrr}
\hline
& \multicolumn{3}{c}{mean ratio to optimum} & ratio to & avg. & avg. \\
Method & $n=100$ & $n=200$ & $n=300$ & lower bound & span & time (s) \\
\hline
ILP (exact)     & 1.000 & 1.000 & 1.000 & --- & 6{,}622.9 & 21.49 \\
\textbf{ML (model)} & \textbf{1.132} & \textbf{1.147} & \textbf{1.143} & \textbf{1.141} & \textbf{7{,}566.2} & 0.31 \\
ML (hybrid)  & 1.740 & 1.648 & 1.667 & 1.672 & 9{,}407.1 & 0.38 \\
ML (greedy)  & 2.428 & 2.419 & 2.405 & 2.384 & 12{,}882.6 & 0.39 \\
k-Inner         & 2.443 & 2.480 & 2.408 & 2.415 & 13{,}892.1 & 0.17 \\
k-Baseline        & 3.370 & 3.394 & 3.311 & 3.302 & 17{,}364.7 & 0.11 \\
k-Budget        & 6.925 & 7.116 & 7.045 & 6.909 & 35{,}758.7 & 5.50 \\
\hline
\end{tabular}
\end{table}

\subsubsection{Very large instances}
On the very large split, with up to $5{,}000$ vertices, $25$ times more than the largest training instance, and up to about $450{,}000$ temporal edges, the exact solver is no longer a usable reference. With a time limit of $120$\,s it proves the optimum on $45$, $29$, $16$ and $0$ of the $50$ instances for $n = 500$, $1{,}000$, $2{,}000$ and $5{,}000$, and at $n = 5{,}000$ it finds no positive lower bound. Table~\ref{tab:xl} reports average spans and running times. The model assignment returns a smaller span than each of the three heuristics on all $200$ instances; its mean span reduction with respect to k-Inner is between $45.1\%$ and $48.9\%$ at every size ($47.2\%$, $[43.5\%, 50.8\%]$, at $n = 5{,}000$), and between $43.5\%$ and $48.9\%$ for each of the three trained networks. On the instances with a proven optimum its mean empirical ratio is $1.157$, $1.148$ and $1.158$ for $n = 500$, $1{,}000$ and $2{,}000$, as on the smaller splits; the proven instances are, however, the easier ones (at $n = 2{,}000$, all have $k \le 5$). Within the same $120$\,s, the solver returns a solution smaller than that of the model assignment on $47$ of the $50$ instances with $n = 500$, but on $18$ of the $50$ instances with $n = 2{,}000$ and on none with $n = 5{,}000$, where its solutions are on average $12.0$ times larger. The learned pipeline needs $15.5$\,s per instance at $n = 5{,}000$, of which $13.6$\,s for the forward pass; at inference the Transformer is applied to blocks of $256$ vertices, which bounds the memory without changing the output.

Two further experiments combine the solver with the learned pipeline. First, we give the solver, on each instance, a time limit equal to the running time of the learned pipeline with the model assignment, not counting the construction of the model, which favours the solver. It then proves no optimum, and its solutions are about ten times larger than those of the model assignment at every size (median ratio between $9.0$ and $10.9$). Second, we pass the solution of the model assignment to the solver as a starting solution, with a time limit of $120$\,s. The solver improves the starting solution on all $200$ instances, by $13.2\%$, $9.9\%$, $5.4\%$ and $1.4\%$ on average for $n = 500$, $1{,}000$, $2{,}000$ and $5{,}000$. Compared with the run without a starting solution, its final solution is smaller on $34$ of the $50$ instances with $n = 2{,}000$ and on all $50$ with $n = 5{,}000$, where it also proves the optimum on $3$ instances, all with $k = 2$. The learned pipeline is therefore also useful as a source of starting solutions for the exact solver on the instances where the solver alone does not find good solutions.

\begin{table}[h!]
\centering
\caption{Very large social test split ($50$ instances per size, $k \in \{2,\dots,10\}$): average span (in thousands) and average time per instance (s). ILP: best solution returned within $120$\,s; its time includes the construction of the model. In parentheses, the number of instances on which the solver proves optimality.}
\label{tab:xl}
\footnotesize
\setlength{\tabcolsep}{4pt}
\begin{tabular}{lrrrrrrrr}
\hline
& \multicolumn{4}{c}{avg.\ span ($\times 10^3$)} & \multicolumn{4}{c}{avg.\ time (s)} \\
Method & $500$ & $1{,}000$ & $2{,}000$ & $5{,}000$ & $500$ & $1{,}000$ & $2{,}000$ & $5{,}000$ \\
\hline
ILP ($120$\,s) & 19.9 (45) & 103.4 (29) & 436.7 (16) & --- (0) & 42.6 & 86.2 & 120.9 & 215.2 \\
\textbf{ML (model)} & \textbf{16.6} & \textbf{35.1} & \textbf{78.1} & \textbf{170.8} & 0.39 & 1.03 & 3.40 & 15.5 \\
ML (hybrid)  & 21.1 & 43.7 & 92.9 & 219.9 & 0.53 & 1.37 & 4.21 & 32.3 \\
ML (greedy)  & 28.4 & 59.5 & 125.6 & 289.4 & 0.54 & 1.39 & 4.27 & 17.7 \\
k-Inner      & 31.2 & 64.8 & 136.1 & 309.0 & 0.58 & 1.31 & 3.10 & 10.6 \\
k-Baseline   & 39.7 & 82.4 & 168.5 & 394.6 & 0.67 & 2.77 & 12.4 & 151 \\
k-Budget     & 83.3 & 172.4 & 349.2 & 823.9 & 15.1 & 32.1 & 72.0 & 278 \\
\hline
\end{tabular}
\end{table}

\subsection{Real-world validation: a high-school contact network}
\label{sec:exp-highschool}
Finally we apply the pipeline, without retraining, to a real face-to-face contact
network: the SocioPatterns high-school deployment of
2012~\cite{fournet2014contact}, in which $180$ students wore RFID proximity
sensors for seven school days, yielding $45{,}047$ records $(t, i, j)$, each stating that $i$ and $j$ were in contact during a $20$-second interval. Timestamps are binned to
$5$-minute resolution, repeated records of the same pair within a bin are merged into one temporal edge, and the log is sliced into one instance per school
day ($145$--$158$ active students and $118$--$137$ distinct time bins per
day), so that one span unit equals five minutes. The data set carries no interval budget, so each day is evaluated at
$k \in \{1, 2, 3, 4, 5, 8, 10\}$, giving $49$ instances. The exact solver,
with a time limit of $120$\,s, proves the optimum on $47$ of them (mean
solve time $11.8$\,s); all proven optima are positive. In this setting a solution gives each student at most $k$ periods, of minimum total duration, such that every recorded contact falls inside a period of at least one of its two participants. On this network the choice of formulation of the exact reference matters: the program of Theorem~\ref{thm:ilp} and the earlier formulation that counts each maximal run as one interval differ on $42$ of the $47$ proven instances, the optima of Theorem~\ref{thm:ilp} being $2.1\%$ smaller in total; all ratios below use Theorem~\ref{thm:ilp}.

Table~\ref{tab:highschool} reports the results. We draw four conclusions.

\emph{The pipeline transfers, with a larger gap than on synthetic data.} Trained only on synthetic graphs, the learned pipeline has a mean empirical ratio of $1.408$ ($[1.363, 1.455]$) with the model assignment and $1.412$ ($[1.356, 1.469]$) with the hybrid assignment, with worst cases $1.78$ and $1.87$; over the three trained networks, the mean ratio of the model assignment is $1.426 \pm 0.017$.

\emph{The public k-Inner implementation fails on most real instances.} It returns no solution on $33$ of the $49$ instances: on every instance with $k \ge 4$ and on $5$ of the $21$ instances with $k \le 3$. The failures occur on the many tied timestamps that binning produces; we have not determined whether they are inherent to the algorithm or specific to its implementation, and we count them as failures. On the $14$ proven instances where k-Inner returns a solution, the mean ratios are $1.47$ for k-Baseline, $1.57$ for the model assignment, $1.66$ for the hybrid assignment and $1.70$ for k-Inner; k-Budget reaches $5.63$ over all instances.

\emph{On this network, the learned pipeline and k-Baseline are not separated with confidence.} k-Baseline is the strongest heuristic, with a mean ratio of $1.649$ ($[1.547, 1.750]$), and it is markedly better than every variant of the learned pipeline at $k=2$ ($1.11$ against $1.58$ for the model assignment). The learned pipeline is better for every other $k$, and its total span over all instances is smaller (ratio of totals $1.47$ for the model and $1.52$ for the hybrid assignment, against $1.62$). Compared instance by instance, the model assignment returns a smaller span than k-Baseline on $40$ of the $49$ instances, but its mean span reduction, $8.5\%$, has a confidence interval $[-3.0\%, 17.5\%]$ that contains zero, and the same holds for the hybrid assignment ($6.8\%$, $[-6.9\%, 17.4\%]$) and for each of the three networks.

\emph{The ranking of the assignment rules changes.} On the synthetic distributions the model assignment beats the greedy and hybrid assignments on almost every instance; on the real days it is still better than the greedy assignment (paired mean reduction $4.3\%$, $[1.8\%, 6.7\%]$), but it is not separated from the hybrid assignment ($0.5\%$, $[-1.5\%, 2.5\%]$), which is better for $k \ge 8$. A school day is closer than the synthetic instances to the regime in which span costs are informative, one dense, contiguous block of activity per person, whereas the synthetic distributions, with $k$ scattered intervals per vertex, reward the learned global structure.

\begin{table}[h!]
\centering
\caption{High-school contact network: mean empirical ratio to the
proven optimum, by interval budget $k$ (seven school
days per $k$; $47$ of $49$ optima proven, both unproven instances have $k=1$; span unit $=$ 5 minutes; bold marks the best non-exact
method in each row). In parentheses, the number of proven instances on which k-Inner returns a solution.}
\label{tab:highschool}
\small
\begin{tabular}{lrrrrrrr}
\hline
$k$ & OPT span & model & hybrid & greedy & k-Inner & k-Baseline & k-Budget \\
\hline
1  & 3{,}319.8 & 1.59 & 1.69 & 1.78 & \textbf{1.49} (5) & 1.97 & 3.18 \\
2  & 1{,}772.6 & 1.58 & 1.68 & 1.78 & 1.79 (7) & \textbf{1.11} & 5.48 \\
3  & 1{,}177.7 & \textbf{1.45} & 1.49 & 1.61 & 1.92 (2) & 1.51 & 6.06 \\
4  &    910.9  & \textbf{1.34} & 1.36 & 1.44 & --- (0) & 1.57 & 5.77 \\
5  &    747.7  & \textbf{1.28} & 1.30 & 1.34 & --- (0) & 1.62 & 5.36 \\
8  &    456.4  & 1.32 & \textbf{1.23} & 1.26 & --- (0) & 1.85 & 5.69 \\
10 &    340.7  & 1.34 & \textbf{1.21} & 1.24 & --- (0) & 2.00 & 7.18 \\
\hline
all & 1{,}158.3 & \textbf{1.41} & \textbf{1.41} & 1.48 & 1.70 (14) & 1.65 & 5.63 \\
\hline
\end{tabular}
\end{table}

\subsection{Robustness: temporal resolution and a second network}
\label{sec:exp-robust}
The results above use a single temporal resolution. We therefore repeated the evaluation, with the same network and no retraining, on the 2012 network binned at $15$-minute and $1$-hour resolution, and on a second, larger SocioPatterns deployment in the same high school in December 2013~\cite{ref_mastrandrea2015}, binned at $5$ minutes ($327$ students over five days; $295$--$312$ active students and $60$--$108$ time bins per day; $35$ instances). Table~\ref{tab:robust} reports the results. Ratios of totals are computed on the instances with a proven, positive optimum; the paired comparisons with k-Baseline use all instances.

\begin{table}[h!]
\centering
\caption{Robustness of the learned pipeline on real contact data, without retraining. In parentheses: the range of the number of timestamps per instance, and the number of instances with a proven positive optimum. RoT: ratio of totals to the proven optimum, over those instances. The last two columns compare each assignment rule with k-Baseline (k-B.) over all instances: the number of instances on which the rule returns a smaller span (wins), and the mean relative span reduction, with $95\%$ bootstrap confidence interval, over the instances where k-Baseline has positive span.}
\label{tab:robust}
\footnotesize
\setlength{\tabcolsep}{4pt}
\begin{tabular}{llrrrl}
\hline
Setting & Rule & RoT & k-B.\ RoT & Wins & Reduction [95\% CI] \\
\hline
2012, 5 min    & model  & 1.47 & 1.62 & 40/49 & $8.5\%$ $[-3.0, 17.5]$ \\
($118$--$137$; $47$) & hybrid & 1.52 &      & 39/49 & $6.8\%$ $[-6.9, 17.4]$ \\
               & greedy & 1.60 &      & 35/49 & $2.0\%$ $[-13.0, 13.6]$ \\
\hline
2012, 15 min   & model  & 1.38 & 1.79 & 43/49 & $26.6\%$ $[16.9, 34.7]$ \\
($42$--$48$; $49$) & hybrid & 1.51 &      & 42/49 & $26.6\%$ $[14.8, 36.9]$ \\
               & greedy & 1.59 &      & 41/49 & $23.3\%$ $[10.5, 34.7]$ \\
\hline
2012, 1 h      & model  & 1.48 & 2.07 & 41/49 & $48.6\%$ $[39.0, 58.0]$ \\
($12$--$14$; $31$) & hybrid & 1.50 &      & 41/49 & $53.6\%$ $[43.4, 63.8]$ \\
               & greedy & 1.55 &      & 41/49 & $52.9\%$ $[42.1, 63.5]$ \\
\hline
2013, 5 min    & model  & 1.19 & 1.49 & 33/35 & $16.7\%$ $[13.2, 20.5]$ \\
($60$--$108$; $23$) & hybrid & 1.21 &      & 29/35 & $14.4\%$ $[10.0, 19.0]$ \\
               & greedy & 1.29 &      & 23/35 & $9.7\%$ $[5.0, 14.6]$ \\
\hline
\end{tabular}
\end{table}

On the 2013 network the exact solver proves the optimum on only $17$ of the $35$ instances within $120$\,s, all with $k \ge 4$, and on $23$ within $30$\,min, all with $k \ge 3$; the $12$ instances with $k \le 3$ that remain unproven have gaps between $0.3\%$ and $4.8\%$, and on them the longer run improves the reference solution by up to $5.7\%$. On the $23$ proven instances the model and hybrid assignments have mean empirical ratios $1.181$ ($[1.160, 1.203]$) and $1.192$ ($[1.164, 1.223]$), against $1.542$ ($[1.483, 1.612]$) for k-Baseline; over the three trained networks the mean ratio of the model assignment is $1.191 \pm 0.010$. Over all $35$ instances, the mean ratios to the solver's lower bound are $1.25$, $1.28$ and $1.51$. Here the learned pipeline is clearly better than k-Baseline: the model assignment returns a smaller span on $33$ of the $35$ instances, with a mean reduction of $16.7\%$ ($[13.2\%, 20.5\%]$). The public k-Inner implementation returns a solution on $15$ of the $35$ instances, $3$ of them with a proven optimum.

On the 2012 network, coarser bins make the learned pipeline more, not less, advantageous against k-Baseline. At $15$ minutes all three assignment rules are better than k-Baseline with confidence, the model assignment by $26.6\%$ ($[16.9\%, 34.7\%]$). At $1$ hour, the instances have only $12$--$14$ timestamps, slightly fewer than any training instance ($16$ to $451$), and $18$ of the $49$ optima are zero; the model assignment returns a smaller span than k-Baseline on $41$ of the $49$ instances, with a mean reduction of $48.6\%$ ($[39.0\%, 58.0\%]$), and reaches span $0$ on $10$ of the $18$ instances with a zero optimum, against $13$ for the hybrid and greedy assignments and $4$ for k-Baseline. Over the four settings, no assignment rule is worse than k-Baseline with confidence, and each is better with confidence in three of them, all except the 2012 network at $5$ minutes. The same holds for each of the three trained networks.

\section{Conclusion and Future Works} 
\label{conclusion}

We have considered \MinTCover{}, a problem defined for complex data
summarization, in the setting of $k \ge 1$ that the literature had left
largely unexplored. We have presented an ILP formulation that solves
moderate size instances exactly; an approximation algorithm which, for
every $k$, achieves factor $T-1$ on the instances whose snapshots are
matchings, improving the best known factor of $4(T-1)$ in the case
$k=1$, and which we have complemented with an NP-hardness result for
every fixed $k$ and a resource-augmentation guarantee for the general
problem; a polynomial-time heuristic based on local search; and a
learning-based method extending DLMinTC+ to arbitrary $k$. The
experimental evaluation, on synthetic graphs with social structure and
on a real-world contact network, shows that the learned pipeline stays
within $1.22\times$ of the proven optimum, that this ratio is stable as
the instances grow by an order of magnitude beyond the training
distribution, and that the method transfers to real contact data without
retraining.

Several directions remain open. On the theoretical side, it is not known
whether the dependence on $T$ in our approximation factor is necessary,
a question already open for $k=1$, nor whether the gap in the
resource-augmentation guarantee can be narrowed from $k$ against
$\lfloor k/2 \rfloor$ to $k$ against $k-1$. On the algorithmic side, it
would be interesting to study variants of our local search that extend
the pruning step, for example when more than one vertex containing an
interval of loss equal to $0$ is considered simultaneously. 

\bibliography{biblio}

\begin{appendices}   
\clearpage
\section{Omitted Proofs of Section \ref{sec:ILP}}
\subsection{Proof of Theorem~\ref{thm:ilp}}
\ilp*

\begin{proof}
For a solution of the program and a vertex $v$, let $A_v = \{t \in [T] : x_v^t = 1\}$ be the set of timestamps at which $v$ is active. We first record two facts.

\emph{Fact 1.} For every $A \subseteq [T]$ with indicator vector $x \in \{0,1\}^T$,
\[
x^{1} + \sum_{t=1}^{T-1} \bigl|x^{t+1} - x^{t}\bigr| + x^{T} \;=\; 2\,r(A).
\]
Indeed, extend $x$ by $x^0 = x^{T+1} = 0$. The left-hand side is then the number of indices $t \in \{0,\dots,T\}$ with $x^{t} \ne x^{t+1}$, and each run of $A$ contributes exactly two of them: one where it starts and one where it ends.

\emph{Fact 2.} Let $v \in V$ and $A \subseteq [T]$.
(a) If an activity timeline of $v$ consists of $j$ intervals and its set of active timestamps is $A$, then its span is $|A| - j$, and $r(A) \le j \le \min(k,|A|)$.
(b) Conversely, if $r(A) \le k$, then for every integer $j$ with $r(A) \le j \le \min(k,|A|)$ there is an activity timeline of $v$ with exactly $j$ intervals whose set of active timestamps is $A$.
For (a), an interval $[s,e]$ contains $e-s+1 = \delta([s,e])+1$ timestamps, and the intervals of a timeline are pairwise disjoint, so $|A| = \sum_{i=1}^{j}(\delta(I_{v_i})+1) = S(\mathcal{T}_k(v)) + j$. Each interval is a set of consecutive timestamps contained in $A$, hence it lies inside a single run of $A$, and every run must be covered; therefore $j \ge r(A)$. Finally $j \le k$ by definition, and $j \le |A|$ because every interval contains at least one timestamp.
For (b), let $R_1,\dots,R_{r}$ be the runs of $A$, of lengths $\ell_1,\dots,\ell_{r}$, so that $\sum_i \ell_i = |A|$. Since $r \le j \le |A|$, we can choose integers $1 \le b_i \le \ell_i$ with $\sum_i b_i = j$. Splitting each run $R_i$ into $b_i$ blocks of consecutive timestamps, and taking each block as an interval, gives $j$ pairwise disjoint intervals, ordered as required by the definition of activity timeline, whose union is $A$; as $j \le k$, they form an activity timeline of $v$.

We now prove that the program and \MinTCover{} have the same optimum:
$(\Rightarrow)$ every covering timeline yields a feasible solution of the
same cost, and $(\Leftarrow)$ every feasible solution yields a covering
timeline of no larger cost.

$\Rightarrow$ Let $\mathcal{T}_k$ be an activity timeline that covers $G$, and for each $v \in V$ let $A_v$ be the set of timestamps at which $v$ is active and $j_v$ the number of intervals of $\mathcal{T}_k(v)$. Set $x_v^t = 1$ if and only if $t \in A_v$, $y_v^t = |x_v^{t+1} - x_v^{t}|$, and $z_v = j_v$. Constraint~\eqref{eq-2} holds because every temporal edge $(u,v,t)$ has an endpoint active at $t$, and constraint~\eqref{eq-3} holds by the choice of $y$. By Fact~1 and Fact~2(a), the left-hand side of constraint~\eqref{eq-4} equals $2r(A_v) \le 2j_v \le 2k$. By Fact~2(a), $z_v = j_v \le \min(k, |A_v|)$, so constraint~\eqref{eq-5} and the bound $z_v \le k$ hold. The objective value is $\sum_{v}(|A_v| - j_v) = S(\mathcal{T}_k)$, again by Fact~2(a). Hence the optimum of the program is at most $\OPT_k(G)$.

$\Leftarrow$ Let $(x,y,z)$ be a feasible solution. By constraint~\eqref{eq-3}, $y_v^t \ge |x_v^{t+1} - x_v^t|$, so by Fact~1 the left-hand side of constraint~\eqref{eq-4} is at least $2r(A_v)$, and thus $r(A_v) \le k$. Let $j_v = \min(k, |A_v|)$. Since $r(A_v) \le |A_v|$, we have $r(A_v) \le j_v$, and Fact~2(b) yields a timeline of $v$ with $j_v$ intervals and set of active timestamps $A_v$, whose span is $|A_v| - j_v$ by Fact~2(a). By constraint~\eqref{eq-5} and $z_v \le k$, we have $z_v \le j_v$, so this span is at most $|A_v| - z_v$. The timelines obtained in this way cover $G$ by constraint~\eqref{eq-2}, and their sum span is at most the objective value of $(x,y,z)$. Hence $\OPT_k(G)$ is at most the optimum of the program.

The two inequalities give equality of the optima. Moreover, applying the second construction to an optimal solution of the program gives a covering timeline whose sum span is at most the optimum, hence an optimal one. The construction computes the runs of each $A_v$ and splits them, in $O(T)$ time per vertex.
\end{proof}

\section{Omitted Proofs of Section \ref{sec:theory-k}}
\subsection{Proof of Proposition~\ref{prop:no-degree}}
\nodegree*
\begin{proof}
We fix $k\ge 2$ and $h\in \mathbb{N}$ and construct the instance s.t. the \oneMinTCover{} layout is respected. Let $B$ be a copy of the complete bipartite graph $K_{k+1,2k+1}$ with left side $X$ of size $k+1$ and right side $Y$ of size $2k+1$; all vertices is $X$ have degree $2k+1$ and all vertices in $Y$ have degree $k+1$. We claim that $B$ satisfies Hakimi’s condition with one unit to spare on every nonempty vertex set:

\begin{equation}
\label{eq:bead-slack}
|E(S)|\;\le\;k|S|-1
\qquad\text{for every nonempty }S\subseteq V(B).
\end{equation}

We observe that every induced subgraph is a biclique $K_{a,b}$ with $a\le k+1$ and $b\le2k+1$,
$|S|=a+b$ and $S$ induces exactly $a \cdot b$ edges. Replacing in equation \ref{eq:bead-slack} gives $ab\le k(a+b)-1$. If $a=0$ or $b=0$ then $S$ induces no edge at all, while $k(a+b)-1\ge k-1\ge0$ because $S$ is nonempty, and the inequality holds. If $a\ge1$ and $b\ge1$, adding $k^2$ to both sides puts the inequality in the equivalent form $(a-k)(b-k)\le k^2-1$, and there are three cases: If $a\le k$ and $b\le k$, both factors are nonpositive and their product is $(k-a)(k-b)\le(k-1)^2\le k^2-1$. If $a\le k$ and $b\ge k+1$, the two factors have opposite signs and the product is
nonpositive. Finally, if $a=k+1$ --- the only remaining possibility,
since $a\le k+1$ --- the product equals $b-k\le k+1$, and $k+1\le k^2-1$
holds precisely because $k\ge2$. For $k=2$ and $S=V(B)$ the inequality holds with equality, so a slack of one unit is all that can be guaranteed for every $k \ge 2$. 

Let $h'=\max(h,3)$. Take $h'$ vertex-disjoint copies
$B_1,\dots,B_{h'}$ of $B$ and add $h'$ \emph{ring edges}, the $i$-th joining an arbitrary vertex of $B_i$ to an arbitrary vertex of $B_{i+1}$, indices taken modulo $h'$; call the resulting multigraph $R$. Since $h' \ge 3$, the ring edges form a simple cycle on $B_1,\dots,B_{h'}$ viewed as single nodes. We claim $R$ is $k$-orientable. Let $S$ be a nonempty vertex set of $R$ and let $j\ge1$ be the number of copies it meets, i.e.\ the number of indices $i$ with $S\cap B_i\ne\emptyset$. The edges induced by $S$ are of two kinds. Those lying inside a copy number at most $k|S\cap B_i|-1$ in each met copy by \eqref{eq:bead-slack}, and summing over the $j$ met copies bounds them by $k|S|-j$. A ring edge, on the other hand, is
induced by $S$ only if \emph{both} of its endpoints lie in $S$, hence
only if both of the copies it joins are met; as the ring edges form a
cycle on the $h'$ copies, and any $j$ nodes of a cycle induce at most $j$
of its edges, there are at most $j$ such ring edges. Adding the two
bounds,
\[
|E(S)|\;\le\;\bigl(k|S|-j\bigr)+j\;=\;k|S| ,
\]
so Hakimi's condition holds and $R$ is $k$-orientable. The single unit of
slack per copy granted by \eqref{eq:bead-slack} is thus exactly what pays
for the ring edges.

Add a \emph{pocket}: two new vertices $p$ and $q$, joined by $2k+1$ parallel temporal edges at the timestamps
$2,3,\dots,2k+2$, together with a single edge between $p$ and an
arbitrary vertex of $B_1$ at timestamp $1$. Give the edges of $R$ pairwise distinct timestamps, all larger than $2k+2$. All timestamps of the resulting temporal graph $G_h$ are distinct, so $G_h$ is an instance of \oneMinTCover{}. Let $R^{+}$ denote the subgraph consisting of $R$ together with the timestamp-$1$ edge at $p$.

It remains to verify the properties claimed in the statement. The instance is connected: each copy $B_i$ is connected, the ring edges join the copies into a single component, $p$ is attached to it by the timestamp-$1$ edge, and $q$ is attached to $p$ by the pocket edges; hence $G_h$ is connected. Every vertex of $Y$ in a copy has global degree at least $k+1$ and every vertex of $X$ at least $2k+1$, since ring edges only increase degrees, while $p$ has global degree $2k+2$ and $q$ has $2k+1$. Hence the minimum global degree is at least $k+1$, and the $(k+1)h' \ge (k+1)h > h$ vertices of the sets $X$ have global degree at least $2k+1$.

Since $G_h$ is connected, it has a single component, and this component is not coverable with span $0$: the set $\{p,q\}$ induces $2k+1 > 2k = k\,|\{p,q\}|$ edges of the union multigraph, which is therefore not $k$-orientable, and Proposition~\ref{prop:span0} gives $\OPT_k(G_h) > 0$.

Finally, we show that $\OPT_k(G_h) \le 2$. The multigraph $R^{+}$ is $k$-orientable: for a vertex set $S$ with $p \notin S$ this was shown above, and for $S \ni p$, since $p$ is incident to exactly one edge of $R^{+}$, we have $|E(S)| \le |E(S\setminus\{p\})| + 1 \le k(|S|-1)+1 \le k|S|$. By Proposition~\ref{prop:span0} there is a timeline of span $0$ covering every edge of $R^{+}$, in which every vertex uses at most $k$ intervals, all singletons; since $p$ is incident to a single edge of $R^{+}$, we may assume that $p$ uses exactly the interval $[1,1]$. Replace it by $[1,2]$ and give $p$ the further $k-1$ intervals $[3,3],\dots,[k+1,k+1]$; give $q$ the interval $[k+2,k+3]$ together with the $k-1$ intervals $[k+4,k+4],\dots,[2k+2,2k+2]$. Each of $p$ and $q$ now uses exactly $k$ pairwise disjoint intervals, $p$ is active at $1,\dots,k+1$ and $q$ at $k+2,\dots,2k+2$, so every pocket edge and the timestamp-$1$ edge are covered. The timelines of all other vertices are unchanged, and the edges of $R$ remain covered because their timestamps exceed $2k+2$. Only the intervals $[1,2]$ and $[k+2,k+3]$ have positive span, so $\OPT_k(G_h) \le 2$.
\end{proof}
\end{appendices}

\subsection{Proof of Theorem~\ref{thm:augment}}
\augment*

\begin{proof}
Solve $\mathrm{LP}(\theta k)$, set $S_v=\{t\in[T]:x_v^t\ge\theta\}$ and give $v$ a timeline of minimum span covering $S_v$. Every temporal edge has $\max(x_u^t,x_v^t)\ge\tfrac12\ge\theta$, so it is covered. From $|S_v|\le\tfrac1\theta\sum_t x_v^t$ and Lemma~\ref{lemma:bridge}(a), whenever $|S_v|>k$,
\[
\mathrm{minSpan}(S_v,k)\le(T-1)\Bigl(\tfrac1\theta\sum_t x_v^t-k\Bigr)
=\tfrac1\theta(T-1)\Bigl(\sum_t x_v^t-\theta k\Bigr)
\le\tfrac1\theta(T-1)\,z_v,
\]
while $\mathrm{minSpan}(S_v,k)=0$ otherwise; summing gives $\ALG\le\tfrac1\theta(T-1)\,\mathrm{LP}(\theta k)$. Finally, the
indicator vector of an optimal timeline with $\lfloor\theta k\rfloor$ intervals per vertex is feasible for $\mathrm{LP}(\theta k)$ with objective at most $\OPT_{\lfloor\theta k\rfloor}(G)$, since by Lemma~\ref{lemma:bridge}(b) its number of active timestamps at $v$ is at most $S(\mathcal{T}_{\lfloor\theta k\rfloor}(v))+\lfloor\theta k\rfloor \le S(\mathcal{T}_{\lfloor\theta k\rfloor}(v))+\theta k$.
\end{proof}

\section{Experiments}
\subsection{Detailed View of the Experiments Setup}\label{app:experiments-setup}
\paragraph*{Synthetic instance generation.} 
The problem is motivated by the reconstruction of activity timelines from
records of human interaction, so the topology underlying our instances
should reproduce the two structural hallmarks that such networks
consistently exhibit: a heavy-tailed degree distribution and a clustering
coefficient far above the random-graph baseline. We therefore generate base
graphs with the Holme--Kim model~\cite{holme2002growing}, which couples
exactly these two mechanisms in a single growth process. Each arriving
vertex attaches its first edge by preferential attachment---a proxy for
popularity, or cumulative advantage in the accumulation of contacts---and
forms each subsequent edge either by a further preferential step or, with
probability $p$, by a \emph{triad formation} step that links it to a
neighbour of the vertex it has just attached to, i.e.\ an introduction
through a mutual acquaintance. The second mechanism is what distinguishes
the generator from plain preferential attachment: it induces triadic
closure, and hence a clustering coefficient that does not decay with $n$,
while leaving the power-law degree exponent intact. We use $m = 3$ and
triad probability $p = 0.5$.

Both properties bear directly on the difficulty of the instances rather
than on realism alone. Degree heterogeneity means the number of temporal
edges incident to a vertex varies over orders of magnitude, so the
$k$-interval budget is slack at the periphery and binding at the hubs,
which are precisely the vertices that determine the objective. Triadic
closure means neighbourhoods overlap, so the covering decisions of adjacent
vertices are locally coupled rather than independent---the regime in which
a purely local, greedy assignment is expected to be brittle.
Table~\ref{tab:graph-models} contrasts this generator with a configuration
model whose degrees are drawn uniformly from $[1,5]$, which serves as a
control with comparable sparsity but no transitivity: its clustering
vanishes as $n$ grows, whereas the Holme--Kim graphs retain clustering
above $0.31$ at every size and a maximum degree that grows with $n$, as
expected of a power-law tail. We do not claim the generator is a faithful
simulation of any particular social network---it reproduces neither
assortative mixing nor community structure---but it isolates the two
features that the covering problem is most sensitive to.

\begin{table}[h!]
\centering
\caption{Structural statistics of the base-graph generators (means over
$20$ seeds per size). The social generator maintains high clustering at
every scale, while the configuration model's clustering vanishes with $n$.}
\label{tab:graph-models}
\begin{tabular}{lrrrr}
\hline
Generator & $n$ & mean degree & max degree & clustering \\
\hline
configuration & 30  & 2.79 & 5.0  & 0.045 \\
Holme--Kim    & 30  & 5.36 & 15.8 & 0.456 \\
configuration & 100 & 2.96 & 5.0  & 0.018 \\
Holme--Kim    & 100 & 5.80 & 31.4 & 0.368 \\
configuration & 300 & 2.95 & 5.1  & 0.007 \\
Holme--Kim    & 300 & 5.93 & 60.5 & 0.312 \\
\hline
\end{tabular}
\end{table}

Temporal activity is planted on a \emph{shared} timeline, mirroring real logs in which all participants share one clock. For each instance we draw the number of vertices $n$ uniformly from the sizes of its split, the budget $k$ uniformly from the budgets of its split, and an event length $\ell$ uniformly from $\{5,10\}$, and we build a new Holme--Kim graph from the instance's own random seed, removing isolated vertices. The horizon $L$ is drawn uniformly from $\{\max(225,10\ell),\dots,450\}$. Every vertex receives $k$ disjoint activity intervals of length $\ell-1$, placed uniformly at random in $[0,L]$ by rejection sampling, with at least one inactive step between two intervals of the same vertex. During each of its intervals, the vertex initiates $\ell$ contacts with neighbours chosen uniformly at random: one at each endpoint of the interval and $\ell-2$ at uniformly random real times inside it. Times are rounded down to integers and duplicate triples $(t,u,v)$ are removed. Every temporal edge therefore has an endpoint active at its timestamp, so the planted timeline is a feasible solution, and every instance has at most $L+1 \le 451$ distinct timestamps, which keeps the quadratic self-attention of the temporal Transformer affordable.

This generator plants exactly $k$ intervals per vertex and makes every event initiated by an active vertex, and the network is trained on the same family; the synthetic experiments therefore measure performance within the training distribution. The real contact network of Section~\ref{sec:exp-highschool}, which has no planted structure and on which we do not retrain, is the out-of-distribution test.

We generate $2{,}000$ training and $300$ validation instances with $n \in \{30, 50, 100, 200\}$ and $k \in \{1,\dots,10\}$. Three quarters of them follow the protocol above; in the remaining quarter the horizon $L$ is drawn from $\{15,\dots,60\}$ and the event length from $\{2,5\}$, so that the training data also contain instances with few timestamps, as produced by coarse binning of real logs. The training instances have between $16$ and $451$ distinct timestamps. Every training and validation instance is labelled by an exact solution: we solve the program of Theorem~\ref{thm:ilp} with a time limit of $120$\,s, keep the instance only if the relative gap of the returned solution is at most $0.1\%$, and otherwise discard it and draw a new one, so that no label comes from a heuristic. This filter slightly favours instances that the solver finds easy. The supervision mask is the activity indicator of the returned solution.

The held-out test splits are generated with the protocol above: a \emph{small} split of $200$ instances with $n \in \{30, 50\}$, a \emph{large} split of $200$ instances with $n \in \{100, 200, 300\}$ and up to about $15{,}000$ temporal edges, and a \emph{very large} split of $200$ instances, $50$ for each $n \in \{500, 1000, 2000, 5000\}$, with up to about $450{,}000$ temporal edges; in all three, $k \in \{2,\dots,10\}$. Each instance is generated from its own seed, which is stored with it. No seed of the training or validation data occurs in a test split, and no instance occurs in two splits. The test splits are labelled independently at evaluation time, and their generation-time labels are not used.
\paragraph*{Architecture and training.}
Table~\ref{tab:hyper} lists the architecture and training parameters. The network has $119{,}921$ trainable parameters and is trained in float32 with PyTorch default initialization, one instance per forward pass, gradient accumulation over $8$ instances, and gradient clipping at norm $1$. Checkpoint selection is performed on the downstream objective rather than on the training loss: after every epoch the validation set is decoded end-to-end with the model assignment rule and Algorithm~\ref{alg:reconstruct}, and the checkpoint with the smallest total validation span is retained. The learning-rate schedule and early stopping are driven by the same quantity. We train three networks from different random seeds, each for at most $60$ epochs; a run takes between $2.9$ and $15.0$ hours on the GPU of an Apple M3 Pro laptop. Their selected checkpoints have validation spans $2{,}331.5$, $2{,}360.6$ and $2{,}381.3$. All tables and figures use the network with the smallest validation span, selected by validation span alone; the text reports, for the main quantities, the mean and standard deviation over the three networks.

\begin{table}[h!]
\centering
\caption{Architecture and training parameters of the learned model.}
\label{tab:hyper}
\small
\begin{tabular}{lp{0.68\textwidth}}
\hline
Component & Setting \\
\hline
Budget embedding & $k_{\max}=10$, dimension $d_k=16$ \\
Input projection & linear $17 \to 64$, ReLU \\
GraphSAGE & $K=2$ layers, width $64$, mean aggregation, separate root weight, ReLU \\
Temporal Transformer & $L=2$ layers, $d=64$, $4$ heads, feedforward width $256$, dropout $0.1$, post-norm \\
Prediction head & linear $64 \to 32$, ReLU, linear $32 \to 1$, sigmoid \\
Loss weights & $\lambda_{\mathrm{imit}}=\lambda_{\mathrm{cov}}=1$, $\lambda_{\mathrm{spar}}=0.1$, $\lambda_{\mathrm{trans}}$: $0.1 \to 2$ over $10$ epochs \\
Loss constants & $\varepsilon = 10^{-7}$, margin $\mu = 1$ \\
Optimizer & Adam, learning rate $10^{-3}$, $\beta=(0.9,0.999)$, no weight decay \\
Schedule & learning rate $\times 0.5$ after $15$ epochs without improvement \\
Epochs & at most $60$; checkpoint with the smallest validation span kept \\
Decoders & model, greedy, hybrid ($\beta=1$), Section~\ref{sec:postproc} \\
\hline
\end{tabular}
\end{table}

\end{document}